\documentclass[jou]{apa}
\usepackage{graphicx}
\usepackage{bm}
\usepackage{mathtools}
\usepackage{amsmath}
\usepackage{amsfonts}
\usepackage{breqn}
\usepackage{booktabs} 
\usepackage{xcolor, soul}

\sethlcolor{yellow}
\newcommand{\norm}[2]{\left \lVert #1 \right \rVert_{#2}}

\usepackage{tikz}

\begin{document}


\title{Hierarchical temporal receptive windows and zero-shot \\timescale generalization in biologically constrained \\[0.4ex] scale-invariant deep networks} 


\rightheader{Hierarchical TRWs and generalization to time-rescaling in scale-Invariant deep networks}

\leftheader{Sarkar \& Howard}
\twoauthors{Aakash Sarkar}{Marc W.~Howard }

\twoaffiliations{%
  Department of Psychological and Brain Sciences\\
  Boston University
}%
{%
 Department of Psychological and Brain Sciences\\
 Department of Physics\\
 Boston University
}

\abstract{

Human cognition integrates information across nested timescales. While the cortex exhibits hierarchical Temporal Receptive Windows (TRWs), local circuits often display heterogeneous time constants. To reconcile this, we trained biologically constrained deep networks, based on scale-invariant hippocampal time cells, on a language classification task mimicking the hierarchical structure of language (e.g., \textit{`letters’} forming \textit{`words’}). First, using a feedforward model (SITHCon), we found that a hierarchy of TRWs emerged naturally across layers, despite the network having an identical spectrum of  time constants within layers. We then distilled these inductive priors into a biologically plausible recurrent architecture, SITH-RNN. Training a sequence of architectures ranging from generic RNNs to this restricted subset showed that the scale-invariant SITH-RNN learned faster with orders-of-magnitude fewer parameters, and generalized zero-shot to out-of-distribution timescales. These results suggest the brain employs scale-invariant, sequential priors---coding ``what" happened ``when"---making recurrent networks with such priors particularly well-suited to describe human cognition.

}

\maketitle

\begin{figure*}
\begin{center}
    \includegraphics[width=0.95\textwidth]{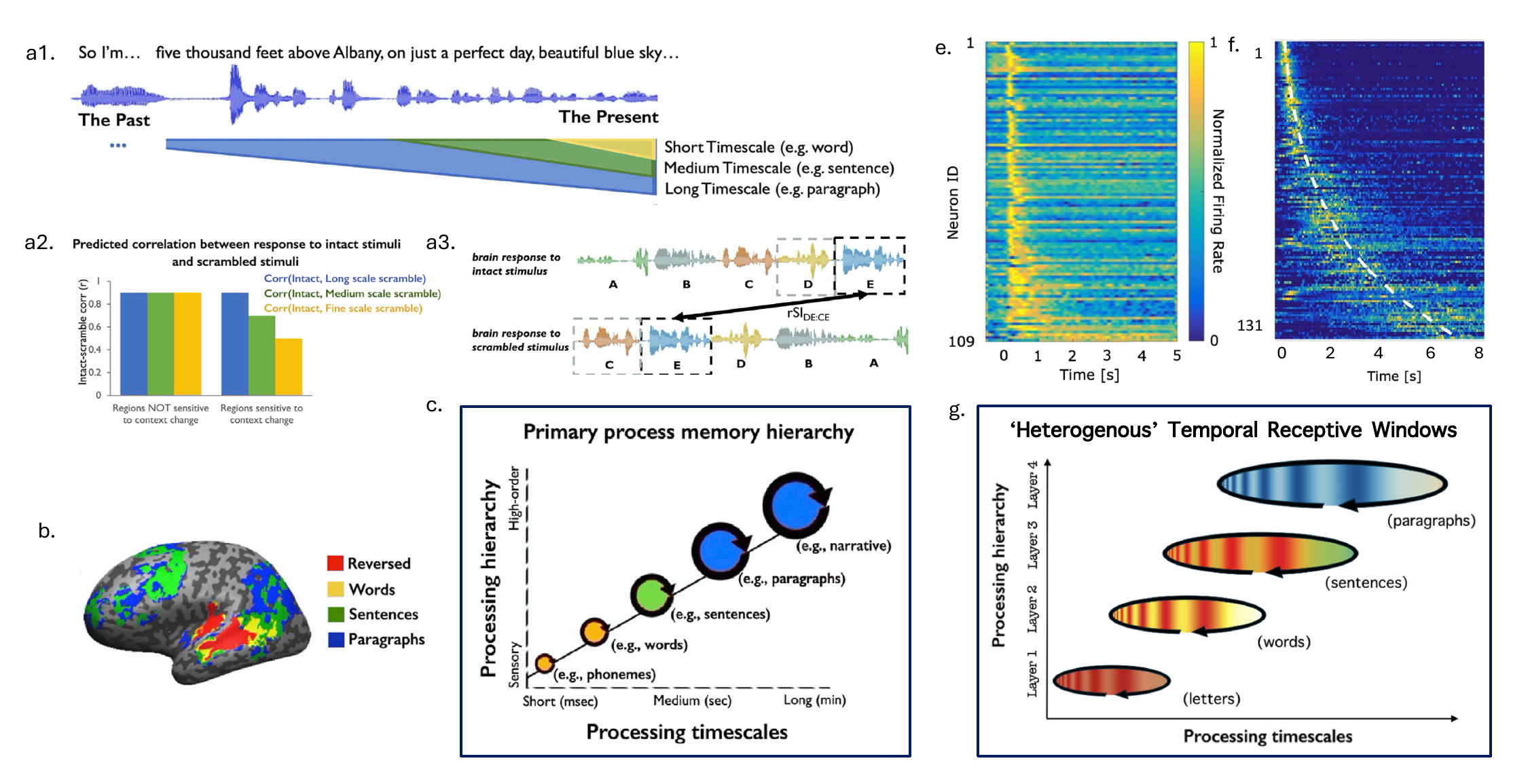}    %
\end{center}
\caption{
\emph{There is evidence to suggest that brain regions, each with a
	heterogeneity of time-scales, can still exhibit a hierarchy of
	increasing Temporal Receptive Windows.} \emph{Left: }Spoken language
	has a multitude of timescales, where one has to keep simultaneously
	keep track of words, sentences and the paragraphs to make sense of the
	current word (\textit{a1}).
Different brain regions need different
	timescales of context to process the current input, defined as its
	Temporal Receptive Window (TRW).
To measure this processing timescale,
	there has been a body of work which measure the correlation of
	activity from a brain region in response to naturalistic stimuli, both
	intact and scrambled (\textit{a2}).
Scrambling the sequence at
	different timescales corresponds to swapping out different lengths of
	context preceding an input (\textit{a3}), and tests how sensitive a
	brain region is to these different scales of context (\textit{b}).
These studies widely support the hypothesis that there is a hierarchy
	of temporal receptive windows in the cortex (\textit{c}), with sensory
	areas processing the immediate context preceding the input, while
	higher-order areas collect and assimilate context from longer
	timescales like sentences and paragraphs.
\emph{Right: } Recent
	studies in memory neuroscience show populations of cells called
	temporal context cells (\textit{e.}) and time cells (\textit{f.})
	seen primarily in the hippocampus and some other brain regions, both of
	which encode a compressed record of the recent past, using different
	temporal receptive fields.
These populations show a continuous
	distribution of single-neuron time constants with more cells coding
	for the recent past.
Together with other converging evidence about a
	heterogeneity of single-neuron timescales within brain regions, these
	raise the question of whether the brain actually has a distribution of
	time constants within each brain region to support flexibility and
	rescaling---however, when processing sequences with a hierarchical
	temporal structure, these brain regions show different emerging
	processing timescales.
Figure \ref{fig:TRWschem}a-c adapted from \protect\citeA{ChieHone20}.
	\label{fig:TRWschem}}
\end{figure*}

Human cognition requires memory over time scales that are long relative to the
intrinsic time scales---membrane time constants, channel kinetics---of
individual neurons.
For instance, in order to understand the meaning of the
end of this sentence, the reader must integrate information from the beginning
of the sentence.
As the narrative structure builds over many sentences and
paragraphs, the time scale over which information must be integrated and
retrieved grows.

Analogous to the way that spatial receptive fields describe the tendency of
neurons in the visual system to respond to information within circumscribed
regions of retinal space, temporal receptive windows (TRWs) define the
duration over which a neural population processes information.
A large body
of work using cognitive neuroimaging methodologies has identified a hierarchy
of TRWs across human cortex
\cite<Fig.~\ref{fig:TRWschem}a-c,>{LernEtal11,HoneEtal12,HassEtal15}.
For
instance, participants listened to a story presented normally or scrambled at
different timescales (words, sentences, paragraphs).
If a region had a TRW at
a particular scale, the response of the region would be invariant to
scrambling the input at smaller time scales.
Whereas early sensory areas,
like the auditory cortex, responded reliably to individual words, even when
their order in the narrative was scrambled, ``higher-order'' brain regions,
including areas in the parietal and frontal lobes, showed reliable responses
only when longer durations, for instance phrases or sentences remained intact.

\begin{figure*}
\begin{center}
    \includegraphics[width=1.0\textwidth]
    {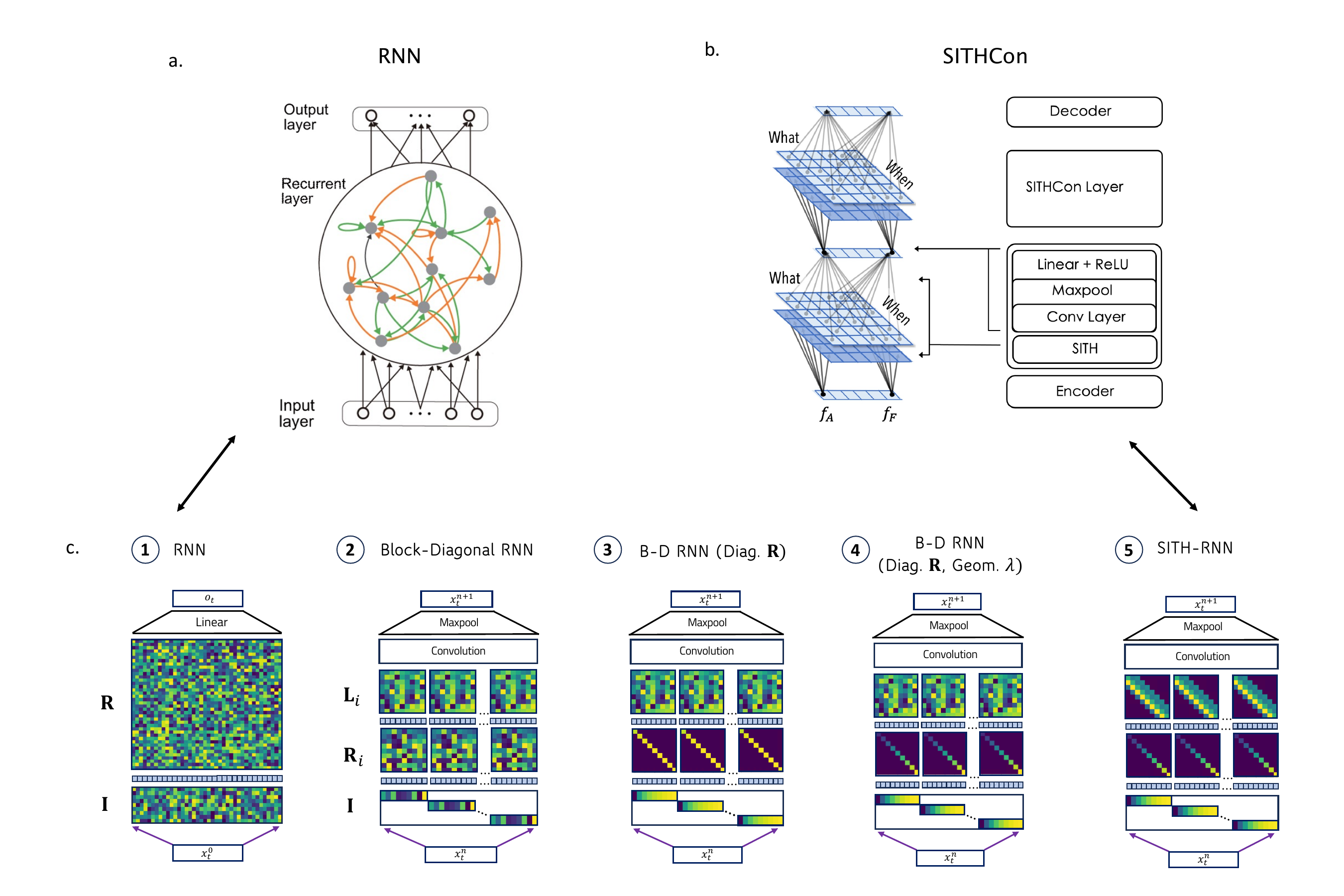}    %
\end{center}
\caption{
\emph{A sequence of RNNs with inductive priors.} \textit{a.} Generic RNNs in neuroscience have
	feed-forward weights that project the input $x_t$ onto the hidden
	layer $(\mathbf{I})$, recurrent weights that operate on the hidden
	layer $(\mathbf{R})$, and feed-forward weights $(\mathbf{L})$ projecting the last
	hidden layer to the output (as shown in \textit{c1.}).
	\textit{b.}~\textit{SITHCon} is a feedforward convolutional neural
	network with inductive priors inspired from populations of time cells
	in the brain, which maintain a log-compressed record of the past.
These priors grant it features like temporal scale-invariance and
	enable it to generalize to slower or faster input without retraining.
Each layer in \textit{SITHCon} has a matrix memory, with time cells
	with different time constants (When) keeping track of every feature
	(What).
At the end of each layer, a convolution catches patterns in
	this log-compressed memory and creates features which can be tracked
	by the next layer.
\textit{c.}~We propose a series of inductive priors
	that successively and gradually builds onto generic RNNs (\textit{c1})
	used in Neuroscience, building up a recurrent network very similar to
	\textit{SITHCon} in \textit{c5.}
To emulate a What x When memory
	structure similar to \textit{SITHCon}, we constrain the projection
	$(\mathbf{I})$ and recurrence matrices $(\mathbf{R})$ to have a block
	diagonal structure (\textit{c2}).
The What x When matrix memory thus
	appears here as a long hidden state, with each feature being tracked
	and evolved independently in its own temporal subspace.
Similar to
	\textit{SITHCon}, a set of output weights $(\mathbf{L})$ generate the
	output state, and a convolution + maxpool layer combines and remaps
	patterns from all features into new features for the next layer.
We
	see that additional priors are required to gain scale-invariant and
	sequential activations, like diagonal Recurrence Matrices (introduced
	in \textit{c3}), geometric eigenvalues in $(\mathbf{R})$ (introduced
	in \textit{c4}), and translated motifs in $(\mathbf{L})$ (introduced
	in \textit{SITH-RNN}, \textit{c5}).
\label{fig:schematic} }
\end{figure*}

\subsection{RNNs as a possible mechanism for hierarchy of TRWs across cortex}

TRWs require that neural population dynamics exhibit memory over time scales
of at least seconds.
Reverberating activity among fast neurons in recurrent
neural networks (RNNs) is a possible cause of slow network dynamics
\cite{Hebb49}.
Computational studies of realistic brain models with recurrent  connections can
show slow network dynamics
\cite{CompEtal00,DingEtal24}.\footnote{It is possible that slow dynamics could also be caused by intrinsic properties
of neurons or synapses
\cite{EgorEtal02,FranEtal05,HassEtal10,ShefEtal11,TigaEtal15,GuoEtal21}.
However  in many cases these networks can be written mathematically as RNNs.
In this paper we do not dwell on the biological causes of these functional
equations.
}
RNNs are mathematically tractable and elegant   
\cite{MaasEtal02,WhitEtal04,RajaAbbo06,DahmEtal19}, making them well-suited
for theoretical analysis.
Moreover, there is a natural relationship between
RNNs and long-lasting patterns of neural activity
\cite{WongEtal07,SussEtal15,RajaEtal16}.

In parallel with this large body of
work in computational neuroscience, work in artificial intelligence has shown
that  RNNs can be trained to perform complex tasks
\cite{Elma90,HochSchm97,GravEtal13,FengEtal24}.

Consider a sequence of neural populations, each with recurrent connections,
that provide input to one another.
Because of recurrent dynamics within each
region, the output of each region ought to be more autocorrelated than its
inputs.
This means that the input to the second layer of the network ought to
be more autocorrelated than the input to the first layer.
Across layers of a
deep RNN, one would expect dynamics to become progressively slower, providing a
possible mechanism for hierarchical TRWs.
To understand deep networks of RNNs, it is essential to understand the network
dynamics within a region.
This paper pursues the implications of a specific
hypothesis for recurrent connections inspired by computational models of human
memory and neural dynamics within regions---hippocampus, entorhinal cortex,
prefrontal regions---implicated in working and episodic memory.

\subsection{Time Cells, Scale-Invariance, and Deep Networks}
In parallel with the hierarchy of time scales across regions, there is
overwhelming evidence for a smooth heterogeneity of functional time constants
\emph{within} brain regions. This basic result  is observed  across species
and tasks
\cite{BernEtal11,WasmEtal18,FascEtal19,CavaEtal20,SpitEtal20,RossEtal19,DansEtal23}
suggesting that within-region
heterogeneity of single-neuron timescales is a fundamental feature of
biological neural networks.
Studies of neural firing patterns in regions believed to be important in
memory suggest that this heterogeneity reflects a more specific form of
temporal coding (Fig.~\ref{fig:TRWschem}e,f).

\subsubsection{Time Cells}

In hippocampus and other regions, so-called time cells fire sequentially after
a triggering stimulus, extending a memory for what happened when over periods
of time extending up to minutes
\cite{PastEtal08,JinEtal09,MacDEtal11,TigaEtal18a,CruzEtal20,TaxiEtal20,ShikEtal21,LiuEtal22,SchoEtal22}.

\subsubsection{Temporal Context Cells}
In addition, so-called temporal context cells have 
receptive fields that decay exponentially in time
with a continuous spectrum of time constants across neurons
\cite{TsaoEtal18,BrigEtal20,NingEtal22,AtanEtal23,ZuoEtal23}. Critically, at least for hippocampal
time cells, the distribution of time
constants appears to be evenly spaced on a logarithmic scale \cite{CaoEtal22}.

This empirical pattern of results---dual populations of temporal context cells
and time cells with logarithmically distributed time constants---can be
identified with a simple
mathematical hypothesis \cite{ShanHowa13}. 
According to this hypothesis, temporal context cells code for the real Laplace
transform of the past as a function of time and time cells code for an
approximate inverse Laplace transform. This mapping provides a high-level
understanding of the information represented by the neural populations which
can thus be used to construct cognitive models of a range of tasks
\cite{HowaEich13,HowaEtal14,HowaEtal18,TigaEtal19,HowaHass20,SaleEtal22}.
and
can be identified in some cases with a continuous attractor neural network
\cite{DaniHowa25,SarkEtal24}.
As we will see, this mathematical hypothesis can be understood as a special
case of an RNN. It is straight forward to write down a linear RNN that gives
out temporal context cells and time cells as a solution \cite{LiuHowa20}. 

\subsubsection{Deep Networks with Time Cells}

\citeA{JacqEtal22} studied the properties of deep networks composed of time cells. In this network (called SITHCon),   time cells are modeled using a matrix memory, which stores the
features (`What') using temporal basis functions with a spectrum of timescales
(`When'). The time-constants are spaced geometrically, which renders the
memory on a logarithmic scale. At the end of each layer, a convolutional layer
searches for patterns on this log-compressed memory, and a maxpool and linear
projection returns the remapped features which would be tracked by the next
layer. 
SITHCon is robust to temporal rescaling, which  stems from the fact that a
rescaling of the input time series results in a translation within its
logarithmically-compressed temporal memory. Convolutional layers are
inherently equivariant to translations, and the subsequent max-pooling
operation makes the network invariant to these translations by discarding the
exact temporal index of the maximum activation. 

\begin{figure*}
\begin{center}
    \includegraphics[width=0.97\textwidth]{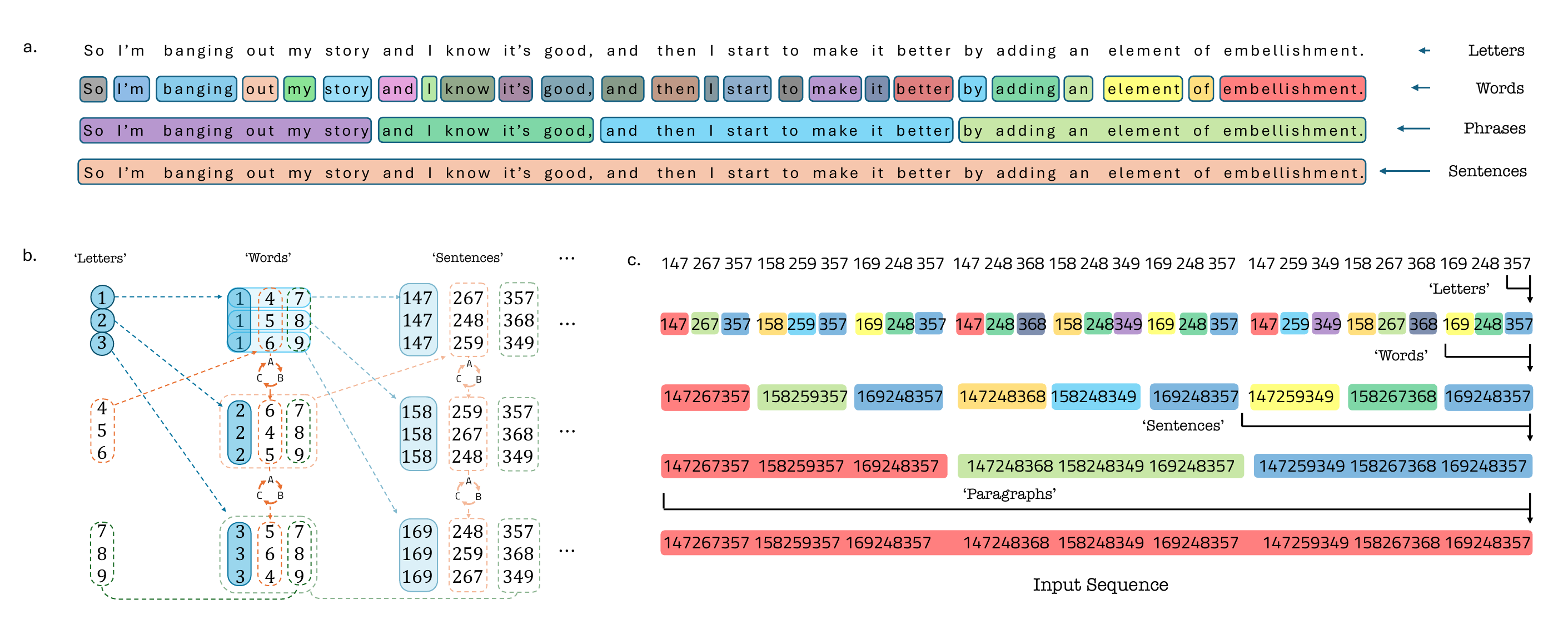}    %
\end{center}
\caption{
\emph{A synthetic hierarchical language with structure across scales.} \textit{a.
}Natural Language has a hierarchy of symbols, with each level combining to
create the next.
We need to keep track of information at these timescales
simultaneously to make sense of the present.
\textit{b.} The toy language is
constructed to have $9$ symbols in each level, with combinations of 3
\textit{`letters'} forming a \textit{`word'}, and so on.
Only certain combinations of symbols are
used, with different rules for each of the three positions---so that knowledge
of the first two symbols is not enough to predict the third symbol. For example, the first \textit{`letter'} in each \textit{`word'} can only be the first three letters ($1,2,$ and $3$), the second positions are permutations of the middle letters ($4,5,$ and $6$), and the third position, and the third position can only be one of the last three \textit{`letters'} ($7,8,$ and $9$), with different rules to populate all possible words. 
The same
motifs are used to generate the symbols from one level to the next---at each
level, the symbol transitions are deterministic.
\textit{c.} The resultant toy
language also has a hierarchy of levels, akin to natural language.
The symbol
transitions at lower levels will still change abruptly when boundaries of
higher-level symbols are reached, and a network attempting to classify such a
hierarchical sequence would thus need to keep track of symbols at multiple
time-scales, akin to maintaining a recollection of the words, paragraphs, and
larger context required to understand speech.
\label{fig:toylang}
} \end{figure*}

\subsection{Overview of the paper}

To investigate the relationship between local scale-invariant dynamics and global hierarchical processing, we adopted a constructive computational approach. We began with defining a behavioral synthetic language task with inherent hierarchical structure across multiple timescales, mirroring the nested composition of natural language. Next, we used a feedforward network model (SITHCon) that explicitly encodes the hypothesized biological constraints---scale-invariance, geometric time constants, and time translation-equivariance---to train on this toy language and observe the emergent processing timescales across layers. Finally, we systematically derived a continuum of recurrent neural networks to determine the minimal set of structural priors required to instantiate these functional properties in a biologically plausible recurrent circuit.

\section{Emergence of a hierarchy of processing timescales}

\begin{figure*}[!th]
\begin{center}
    \includegraphics[width=1.0\textwidth]{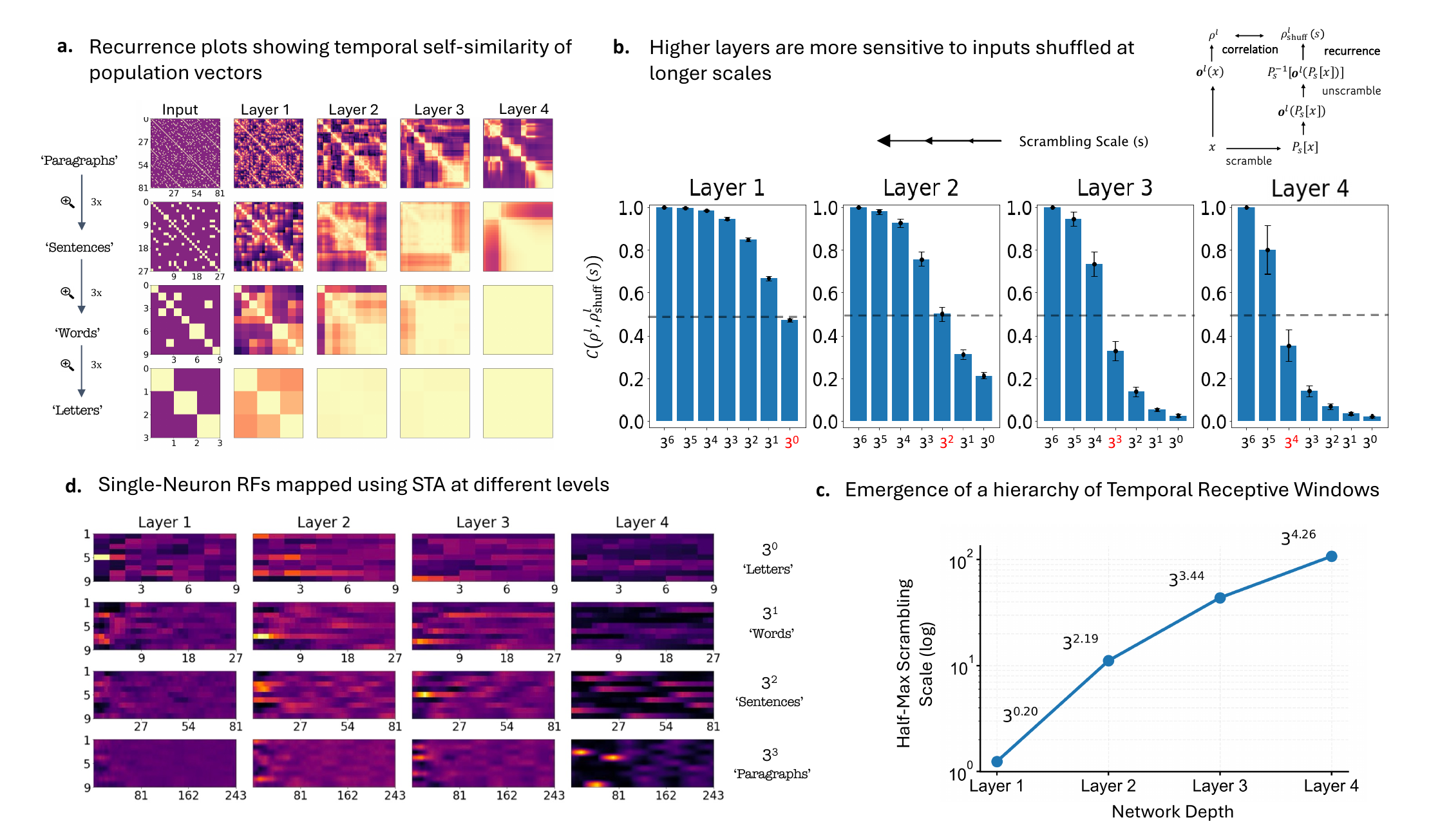}    %
\end{center}
\caption{
\textit{Emergence of hierarchical temporal receptive windows in SITHCon.}
\textit{a. Recurrence plots reveal increasing temporal structure across layers.} Plots show the temporal self-similarity of population vectors for the input (left) through Layer 4 (right). Rows show successive $3\times$ temporal magnifications (top to bottom). Early layers exhibit fine-grained temporal dynamics, while deeper layers maintain stable states over longer durations.
\textit{b. Sensitivity to hierarchical scrambling.} Akin to the methodology in TRW literature, we measured the correlation between responses to original sequences and sequences scrambled at varying timescales (schematic, top right). Early layers are sensitive only to fine-scale scrambling (small $s$) and remain robust to large-scale scrambling (high correlation at large $s$). In contrast, deep layers (e.g., Layer 4) depend on long-range structure and are therefore sensitive to scrambling at both fine and coarse scales (up to $3^4$), recovering only when the scrambling scale exceeds the length of the training sequence ($>3^4$). The grey dashed line indicates the 50\% correlation threshold used to calculate the TRW.
\textit{c. An emergent hierarchy of temporal integration windows.} The effective Temporal Receptive Window (TRW) for each layer, calculated as the half-max scrambling scale ($s_{50}$) from the data in \textit{b} using linear interpolation. The approximately linear trend on the semi-log plot indicates exponential expansion of integration timescales; the slight saturation at Layer 4 reflects the network matching the bounded global timescale of the training sequences ($>3^4$).
\textit{d. Layer-wise tuning to hierarchical levels}. Spike-Triggered Averages (STA) map linear receptive fields of representative neurons (columns) to stimuli at different hierarchical levels (rows). Layer 1 neurons (left) track elementary symbols but ignore higher-order structure. On the other hand, Layer 4 neurons (right) exhibit a nested compositional structure: a single broad activation band at the \textit{`paragraph'} level resolves into three distinct bands at the \textit{`sentence'} level, and so on, exhibiting signs that deep neurons have learned the hierarchical mapping of the toy language, defining their receptive fields through the specific combinatorial syntax of the grammar (for the RFs of all 9 neurons in each layer, refer to Fig.~\ref{fig:STAsupp}).
\label{fig:TRW}
}
\end{figure*}

Before investigating the dynamics of recurrent circuits, we first test the fundamental hypothesis: can a system with identical, log-compressed time constants in every layer spontaneously organize into a processing hierarchy? To answer this, we utilize SITHCon, a feedforward architecture that acts as a \emph{canonical} scale-invariant reference architecture. By explicitly hard-coding a log-compressed memory basis (approximating perfect time cells), SITHCon allows us to isolate the functional consequences of this inductive bias without the added complexity of training recurrent dynamics.

\subsection{Task: A hierarchical toy language}

We develop a hierarchical toy language with symbols corresponding to letters, words, and sentences, with deterministic transitions and different ascribed timescales for each level (Fig.~\ref{fig:toylang}).

Formed with symbols {1-9}, these sequences have fixed combinations which become the next hierarchical unit---3 \textit{`letters'} combine to create \textit{`words'}, 3 \textit{`words'} combine to create \textit{`sentences'}, and so on.
Only certain combinations are allowed, and knowledge of the first two symbols would be vital to predicting the third.
The rules are motivated to create XOR-like combinations---with the first, second, and third positions in the combination always chosen from the first three symbols ($1-3$), middle three symbols ($4-6$), and the last three symbols ($7 -9$) respectively.
This also ensures that in the final sequences, the marginal statistics for each `letter' is uniform and thus they are each equally represented.
Knowledge of symbols in both the first and second positions uniquely determines the third symbol, and thus the identity of that combination.
Since the combinations are happening at different length-scales, to understand the nature of the sequence would necessitate remembering the sequence at the multiple, hierarchical timescales, akin to how one would listen and comprehend speech.

The final constructed sequence consists of 4 specified hierarchical levels, generating sequences with $81$ symbols.
The network's task involves classifying sequences based on different symbol combinations across these levels. We train the networks to classify these sequences at regular scale.

\subsection{Deep layers spontaneously organize distinct temporal event boundaries}

To visualize how the network segments the continuous input stream, we analyzed the temporal self-similarity of the population activity in each layer (Fig.~\ref{fig:TRW}a). We computed recurrence plots based on the Pearson correlation between the population output vectors $\mathbf{o}^l$ at different time points (see Methods). Geometrically, this metric quantifies the cosine of the angle between the z-scored population states at times $t_i$ and $t_j$, allowing us to visualize the stability of the neural representation over time.

In the early layers (Fig.~\ref{fig:TRW}a, Left), the recurrence plots exhibit a fine-grained, checkerboard pattern. This indicates that the population state changes rapidly, effectively resetting at the boundaries of individual symbols to represent the immediate sensory input. However, as we progress to deeper layers (Fig.~\ref{fig:TRW}a, Right), a striking transformation occurs: large, stable blocks of high self-similarity emerge off the diagonal. These blocks indicate that the neural population is maintaining a stable representation over extended durations, effectively ignoring the high-frequency jitter of individual input tokens. The sharp transitions between these stable regions seem to align with the hierarchical boundaries of the task structure (words and sentences). This suggests that despite the absence of explicit segmentation cues during training, the network spontaneously chunks the input into increasingly abstract narrative units, with deeper layers operating on a \textit{renormalized} timeline of events rather than raw time steps.

\subsection{Emergence of hierarchical Temporal Receptive Windows mirroring linguistic structure}

To quantify the effective processing horizon of each layer, we applied a temporal scrambling perturbation analysis mirroring functional neuroimaging studies \cite{LernEtal11}. We permuted the input sequence $x$ at various shuffling scales $s$, ranging from individual symbols ($3^0$) to a maximal length of 9 training sequences (placed side by side, $9*3^4 = 3^6$), and compared the layer's ($l$) response to the scrambled input against its response to the intact sequence, $\mathbf{o}^l(x)$.

Crucially, because scrambling disrupts the timeline, we cannot compare the output sequences directly. Instead, we first ``un-shuffle'' the layer's response to the scrambled input by applying the inverse permutation, $P_s^{-1}[\mathbf{o}^l(P_s[x])]$, thereby realigning it temporally with the original input sequence\footnote{Although the un-shuffled output aligns temporally with the original input, the \textit{context} preceding each time step remains scrambled at scale $s$. To rigorously quantify this discrepancy, we compute the correlation between the recurrence matrix of the original activity $(\mathbf{o}^l(x))$, denoted as $\rho^l$, and that of the un-shuffled activity $P_s^{-1}[\mathbf{o}^l(P_s[x])]$, denoted as $\rho^l_{\tt{shuff}}(s)$ . This metric, $C(\rho^l,\rho^l_{\tt{shuff}}(s))$, measures how much the layer relies on temporal context extending beyond $s$.}.
By observing the discrepancy between the original and un-shuffled activity patterns as a function of $s$ (Fig.~\ref{fig:TRW}b), we can describe the layer's sensitivity to the timescale of prior context.

The results reveal a clear dissociation in temporal sensitivity. Early layers show a drop in correlation only when the input is scrambled at very fine scales (e.g., individual symbols); they remain largely unaffected by coarse scrambling because their integration window is too short to detect the disruption of long-range context. In contrast, deeper layers exhibit a sharp drop in performance even when the input is scrambled at large scales. 

To quantify this scaling, we defined the effective Temporal Receptive Window (TRW) for each layer as the scrambling scale $s_{50}$ where the correlation drops to half-maximum (calculated using linear interpolation, Fig.~\ref{fig:TRW}c). This metric provides a robust, parameter-free estimate of the integration timescale, allowing us to compare the \textit{processing bandwidth} across layers. We observed that these timescales expand exponentially with network depth, indicated by the linear progression on the semi-log plot. Specifically, the integration window scales from elementary symbols in Layer 1 ($s_{50} \approx 3^{0.2} $) to nearly the training sequence length in Layer 4 ($s_{50} \approx 3^{4.3}$), confirming that the network has autonomously organized into a hierarchy of exponentially increasing processing timescales.

\subsection{Neurons in deeper layers develop \textit{`sentence'} and \textit{`paragraph'}-level receptive fields}

Finally, we move from population-level analyses to the specific representational content of individual neurons. We utilized Spike-Triggered Averages (STA) to compute the linear Receptive Field (RF) of neurons—identifying the specific stimulus features most predictive of a neuron's firing. While standard STA typically assumes a single timescale, our hierarchical task allows us to compute the RF at multiple levels simultaneously: we can ask what \textit{`letter'}, \textit{`word'}, or \textit{`sentence'} maximally drives a specific neuron. This approach is particularly valid here, as the simplified SITHCon network (ablating the final nonlinearity) allows for a direct linear interpretation of the neuronal response.

Fig.~\ref{fig:TRW}d displays the linear RFs of representative neurons across the four layers (left to right), constructed at four hierarchical levels (top to bottom). The output of Layer 1 functions as a simple symbol detector, showing discernible activation only at Level 1 (\textit{`letters'}) while remaining blind to higher-order structures. However, as we ascend the hierarchy, neurons begin to exhibit \emph{feature abstraction}. A representative neuron in Layer 4 (far right) does not respond to individual symbols; instead, it is tuned to a specific high-level sequence (a \textit{`sentence'} or \textit{`paragraph'}).

Notably, this high-level selectivity is \emph{compositional}. Extending this analysis to the full population reveals a robust, fractal-like fractionation of receptive fields (Fig.~\ref{fig:STAsupp}): the broad activation band of a Layer 4 neuron at the \textit{`paragraph'} level can be explicitly decomposed into a preference for the specific combination of three `sentences', which can be further decomposed into nine \textit{`words'}, and so on. This demonstrates that the network performs a sort of layer-wise `temporal renormalization': each layer compresses its input history into discrete features that serve as the effective, atomic time-steps for the next layer. This mechanism allows the system to construct an understanding of the narrative structure without requiring any explicit segmentation cues.

\section{Recurrent implementations of scale-invariant dynamics}

The previous section demonstrated that a log-compressed temporal basis distribution (SITHCon) is sufficient to generate hierarchical TRWs functionally. However, to implement this mechanism in biological circuits, it must be realizable within a recurrent dynamic framework. RNNs represent the standard computational tool for modeling neural dynamics in neuroscience, yet generic RNNs lack interpretability and structure. Here, we bridge this gap by distilling the key mathematical properties of SITHCon—scale-invariance, modularity, and translation-equivariance—into a biologically plausible Recurrent Neural Network (SITH-RNN). This allows us to propose SITH-RNN not just as a machine learning architecture, but as a candidate for normative architectural framework for cortical processing.

\subsection{Deriving the SITH-RNN architecture}

\subsubsection{Scale-invariant activity from recurrent dynamics}
Linear recurrent networks have been shown to generate sequential scale-invariant activity under certain constraints of the recurrent matrix \cite{LiuHowa20}.
Scale-invariance of neural activity implies that the responses of any two neurons in the sequence are rescaled versions of each other in time.
Mathematically, this requirement can be written as
\begin{equation}
x_i(t)= x_j(\alpha_{ij}t), \,\, \forall i, j \in 1,2, \cdots,N
\label{eq: scale-inv}
\end{equation}
which means that for every pair of neurons $i$ and $j$, there exists a factor $\alpha_{ij}$, such that  their neural activity, $x_i(t)$ and $x_j(t)$, respectively, are rescaled in time by that factor.

Recent work has shown that single-layer recurrent networks with a continuous-time dynamics on the hidden state $\mathbf{h(t)}$ and a recurrent matrix $\mathbf{R}$

\begin{equation}
\frac{d\,\mathbf{h}(t)}{dt} = \mathbf{R}\mathbf{h}(t)\;, \label{eq:contrnn}
\end{equation}
can maintain scale-invariant activity, subject to certain constraints---the eigendecomposition of $\mathbf{R}$ must yield both geometric eigenvalues, and translation-invariant eigenvectors \cite{LiuHowa20}. Linear networks with these restraints can produce monotonically decaying temporal dynamics in $\mathbf{h}_t$ , and even exhibit scale-invariant sequential activity (like time cells) when the network was extended by a set of downstream linear weights $\mathbf{L}$\footnote{The activity of such networks with two sets of weights could be equivalently represented with a single set of effective recurrent weights, defined as $\mathbf{R}_{\tt{eff}} = \mathbf{L} \mathbf{R} \mathbf{L}^{-1}$---as this form is identical to an eigendecompositon of $\mathbf{R}_{\tt{eff}}$, the constraints mentioned above reduce to $\mathbf{R}$ being diagonal with geometric eigenvalues, and $\mathbf{L}$ having columns with repeated, translated motifs, which supplies the translated eigenvectors.}. This section studies and extends these constraints in deep recurrent neural networks with multiple input features.

\subsubsection{Block-diagonal weight matrices produce a `What' $\times$ `When' representation} A generic RNN has a hidden state $\mathbf{h}_t$, which is updated with new input $x_t$ and generates an output $\mathbf{o}_t$ at each time point.
The evolution of a single layer of a generic RNN can be modeled using a discrete-time version of Eq.~\ref{eq:contrnn}, with feedforward weights $\mathbf{I}$ and $\mathbf{L}$ projecting in and out of the hidden space respectively

\begin{eqnarray}
    \mathbf{h}_{t} = \mathbf{R} \mathbf{h}_{t-1} + \mathbf{Ix}_t\;, \label{eq:rnn} 
    \\ \mathbf{o}_t =  \mathbf{L}\mathbf{h}_t \label{eq:o_t}\;.
\end{eqnarray}
In standard RNN formulations, the input ($\mathbf{I}$) and recurrent ($\mathbf{R}$) weights are dense and fully trainable, resulting in unstructured representations that mix conflate stimulus identity (‘What’) with temporal history (‘When’). 

To prevent this entanglement, we enforce a structural prior that preserves independent timelines for each feature \cite{SarkEtal24}, by defining the hidden state as a tensor product of feature and temporal subspaces

\begin{dmath}
    \mathbf{h_t} = \bigotimes^{n_f}_{k=1}\mathbf{h_t^k} \label{eq:prod-hid}
\end{dmath}
where $\bigotimes$ represents a tensor product, $n_f$ is the number of features, and $\mathbf{h^k_t}$ is the temporal subspace for each feature $k$. Each temporal subspace $\mathbf{h^k_t}$ consists of $n_{\tau}$ neurons with different time constants to create a memory of that feature\footnote{This renders $\mathbf{h_t}\in \mathbb{R}^{n_{f} \times n_{\tau}}$, reminiscent of the `What' $\times$ `When' matrix memory used in SITHCon, albeit with the difference that here the temporal subspaces $\mathbf{h^k_t}$ can be concatenated and represented as a vector.}.

The recurrent ($\mathbf{R}$) and feedforward weights ($\mathbf{I}$) must also possess additional structure to preserve this tensor product structure of `What' and `When' in 
$\mathbf{h_t}$, and ensure that the temporal sub-spaces for each feature evolve independently---namely, they must have separate operators $\mathbf{R}_k$ acting on each subspace, implying a block-diagonal structure $\mathbf{R} = \bigotimes^N_{k=1}\mathbf{R}_k$. Additionally, each feature must evolve identically (to maintain separable representation) and hence the matrices should have identical sub-blocks, or equivalently, be a tensor product of corresponding spaces for feature and time 

\begin{eqnarray}
    \mathbf{R} = \mathbf{R}_{what} \otimes \mathbf{R}_{when} \label{eq:prod-R}
\end{eqnarray}
where for our purposes, $\mathbf{R}_{what}$ is a diagonal matrix as the feature space is assumed to be orthogonal, with similar priors on $\mathbf{L}$ and $\mathbf{I}$ \footnote{While the entire recurrent matrix $\mathbf{R}$ has $(n_f \times n_{\tau})^2$ total entries, the actual effective parameters are only for the block $\mathbf{R}_{when}\in\mathbb{R}^{n_\tau \times n_{\tau}}$, which is repeated for each feature. Similarly for the projection weights, we have $\mathbf{L}_{when}\in\mathbb{R}^{n_\tau \times n_{\tau}}$, and  $\mathbf{I}_{when}\in\mathbb{R}^{n_f \times n_{\tau}}$.}.

Often the task at hand might require a remapping of features from layer to layer, (called \textit{channel mixing} in contemporary ML architectures, like state-space models)---to do this, we introduce a convolution layer (analogous to the convolutional layer in SITHCon which makes it robust to temporal rescaling) after the downstream weights $\mathbf{L}$. Thus, for our case, Eq.~\ref{eq:o_t} is updated to

\begin{eqnarray}
    \mathbf{o}_t =  \max_{n_{\tau}} (g \star \mathbf{L}\mathbf{h}_t) \label{eq:conv-L}
\end{eqnarray}

where 2-d convolutional filters $g$ searches for patterns on a feature $\times$ time matrix-reshaped version of  $\mathbf{L}\mathbf{h}_t$ (which allows the kernel to see the history of all features simultaneously, instead of the regular concatenated vectorized form), and a max-pool operation ($\max_{n}$) which runs over the temporal dimension returning a trace for each pattern, returning a combination of detected features for the next layer (for a detailed schematic see Fig.~\ref{fig:sithrnn}).

\subsubsection{A continuum of constraints: From generic RNNs to SITH-RNN}
To isolate the contribution of the specific constraints derived above—block-diagonal structure, fixed geometric timescales, and convolutional readouts—we constructed an architectural continuum of five networks (Fig.~\ref{fig:schematic}, also reproduced with their temporal dynmamics and generalization performance in Fig.~\ref{fig:seqRNN}). 
While \textit{Network 1} is a standard linear system, \textit{Networks 2 through 5} share the tensor product structure described in Eqs.~\ref{eq:prod-hid}--\ref{eq:conv-L}, using a shared convolutional feature mixer ($g$) and recurrent blocks ($\mathbf{R}_{when}$) that operate independently on temporal subspaces.

\begin{description}
    \item[Network 1 (Generic RNN)] serves as the baseline linear dynamical system where $\mathbf{I}$ and $\mathbf{R}$ are dense and fully trainable.
    
    \item[Network 2 (Dense Block-Diagonal)] enforces the tensor product structure but retains dense, trainable recurrent sub-blocks ($\mathbf{R}_{when}$) and projection weights ($\mathbf{I}_{when}$ and $\mathbf{L}_{when}$), allowing for arbitrary mixing within the temporal subspace.
    
    \item[Networks 3 \& 4 (Diagonal RNNs)] constrain the recurrent matrix $\mathbf{R}_{when}$ to be diagonal and fix the input projection to $\mathbf{I}_{when} = \begin{bmatrix} 1 \cdots 1 \end{bmatrix}$ to project identically to all neurons in the temporal space of each feature. This forces the hidden state to act as a bank of leaky integrators with different time constants.
    \begin{itemize}
        \item \textit{Network 3} utilizes \textit{uniformly distributed} time constants.
        \item \textit{Network 4} utilizes \textit{geometrically distributed} time constants.\footnote{This introduces the first constraint for scale invariance: if the recurrent and feedforward dynamics are viewed as a single effective operator $\mathbf{R}_{\tt{eff}}$, this step ensures $\mathbf{R}_{\tt{eff}}$ possesses geometric eigenvalues.}
    \end{itemize}
    
    \item[Network 5 (SITH-RNN)] completes the transition by fully decoupling temporal memory from feature binding. Unlike the previous networks where $\mathbf{L}_{when}$ was fluid, SITH-RNN freezes the recurrent connectivity (using the geometric eigenvalues from Network 4) and restricts learning in $\mathbf{L}_{when}$ to a translated \textit{temporal motif} of fixed width, which is implemented as a banded (Toeplitz) matrix.\footnote{This introduces the second constraint for scale invariance: the banded structure implies that the eigenvectors of the effective recurrent matrix $\mathbf{R}_{\tt{eff}}$ are translation-invariant \cite{LiuHowa20}.}
\end{description}

\begin{figure}[!th]
\begin{center}
    \includegraphics[width=0.7\columnwidth]{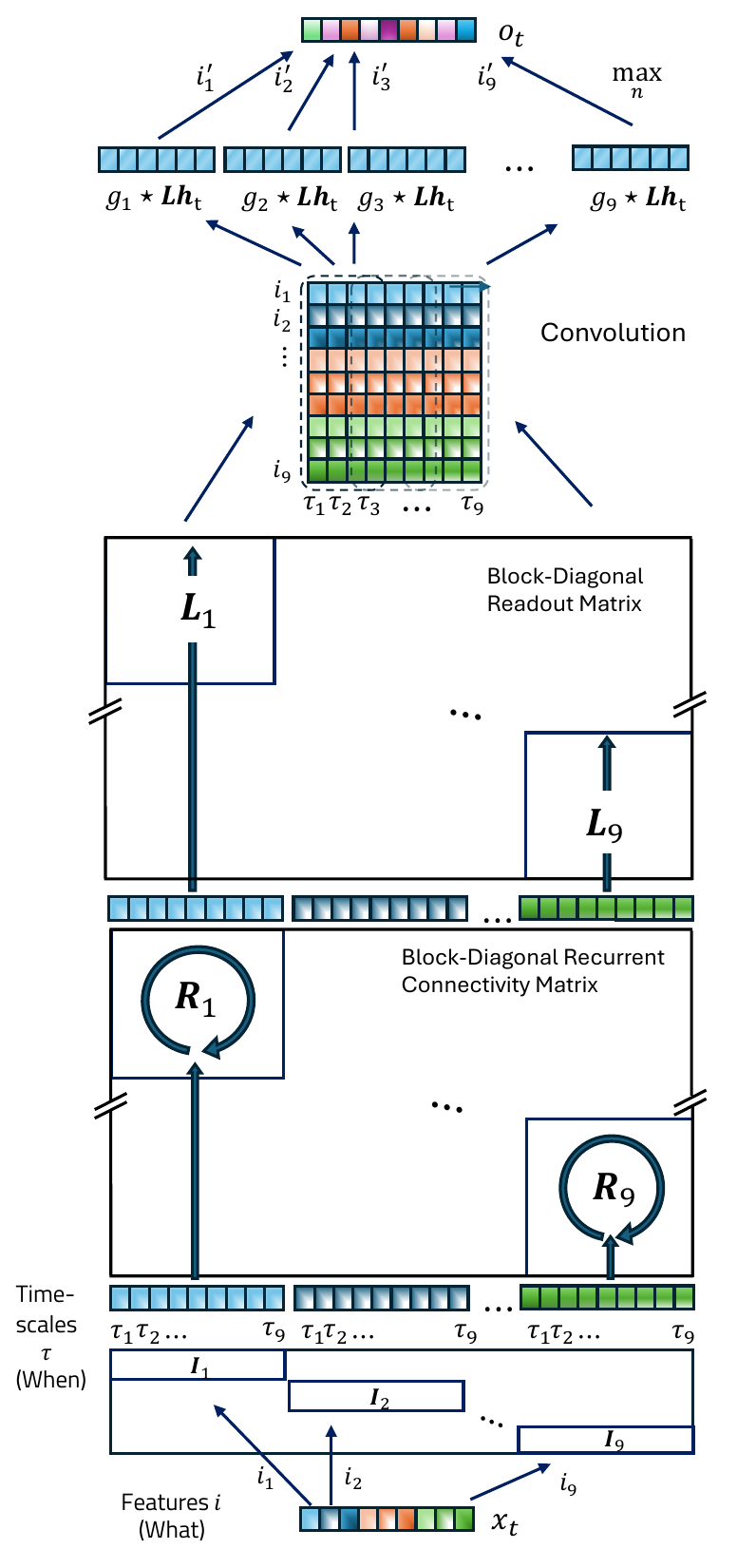}    
\end{center}
\caption[A Schematic of SITH-RNN.]{\emph{A Schematic of SITH-RNN.} We examine a series of RNNs with block-diagonal weights, which retains a compressed memory for each feature.
A Projection Matrix with identical blocks $\mathbf{I}_{when}$ creates a hidden state with separate memory (When) for each feature.
The memory for each feature is evolved independently, using recurrent blocks $\mathbf{R}_{when}$ and readout blocks $\mathbf{L}_{when}$ for each feature.
After stacking the temporal subspaces back into a `What' $\times$ `When' matrix, a convolution tracks pattern, and a maxpool operation collapses across the time dimension to produce the remapped features for the next layer.
\label{fig:sithrnn}}
\end{figure}

The combination of fixed geometric dynamics and translation-invariant readout kernels ensures that SITH-RNN is scale-invariant by construction, searching for the same relative patterns across all timescales simultaneously (for a detailed schematic of the tensor operations and signal flow within SITH-RNN, see Fig.~\ref{fig:sithrnn} in Methods).

\subsection{Scale-invariant time cells emerge from geometric recurrent eigenvalues and translated readout eigenvectors}

\begin{figure*}[!ht]
\centering
    \includegraphics[width=0.99\textwidth]{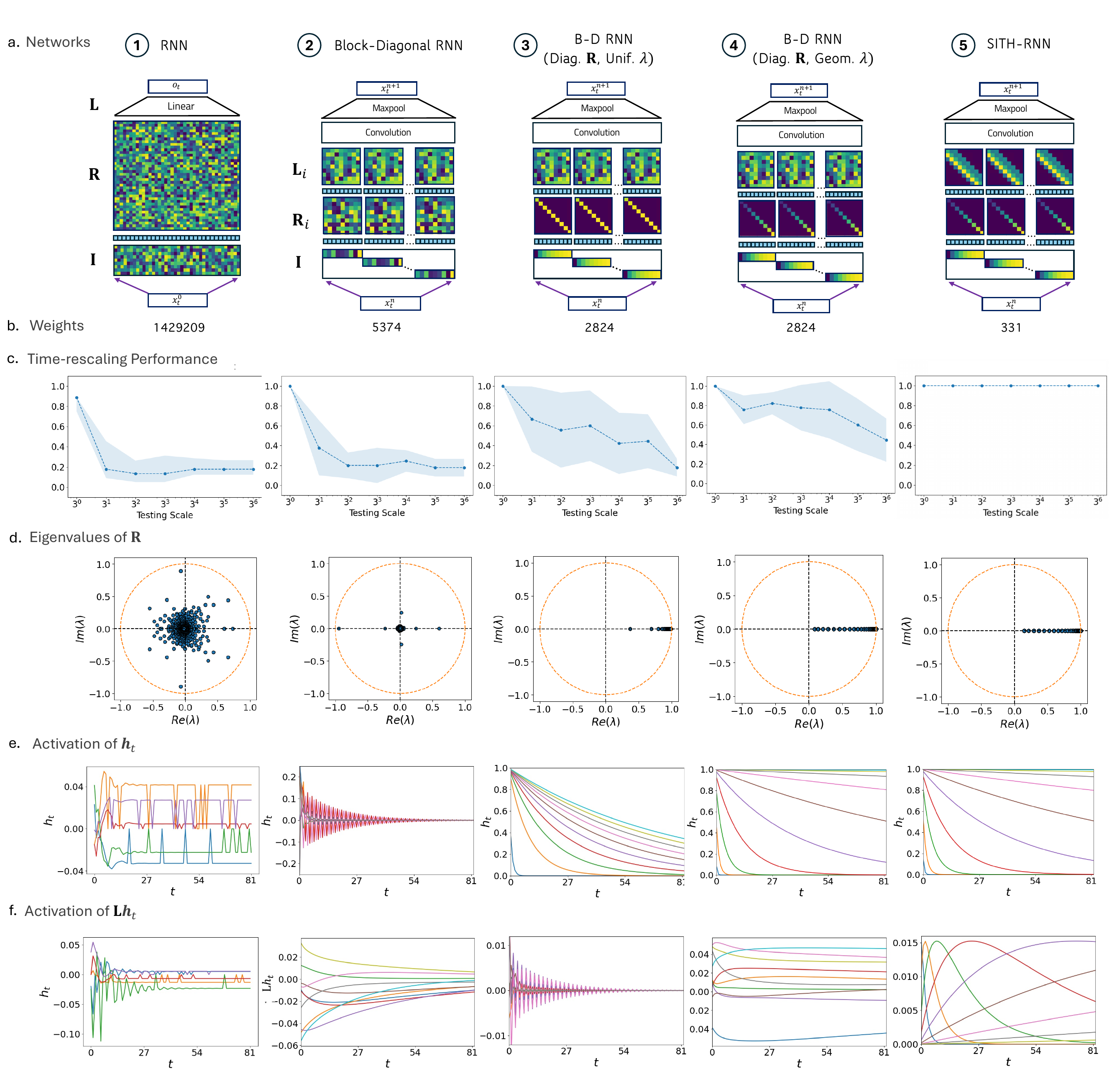}    
\caption{
\textit{Biologically constrained SITH-RNNs achieve superior generalization with fewer parameters by enforcing scale-invariant dynamics.}
\textit{a. Evolution of architectural constraints.} We derived a continuum of five recurrent networks, starting from a generic linear RNN (\textit{Left}) and systematically adding inductive priors motivated by the SITHCon architecture (e.g., block-diagonal connectivity, geometric time constants), culminating in SITH-RNN (\textit{Right}).
\textit{b. Parameter efficiency}. The addition of these structural constraints dramatically reduces the number of trainable weights (by orders of magnitude) compared to the generic RNN, despite identical hidden state dimensions.
\textit{c. Zero-shot generalization to time-rescaling.} When trained on hierarchical sequences at a fixed timescale ($3^0$) and tested on sequences rescaled by factors up to $3^6$, models perform increasingly well as priors are added (Left to Right). SITH-RNN achieves perfect classification accuracy across all six orders of magnitude of timescale, demonstrating robust zero-shot generalization.
\textit{d. Spectral structure of recurrence.} The eigenvalues of the recurrent matrix $\mathbf{R}$ transition from a uniform distribution on the complex unit circle (Generic RNN, \textit{Left}) to real values localized on the axes (Block-Diagonal) and finally to a geometric spacing that tiles the real axis (SITH-RNN, \textit{Right}), a necessary condition for logarithmic (Weber-Fechner) compression and scale-invariance.
\textit{e. Temporal dynamics of the hidden state ($\mathbf{h}_t$)}. Response of the first-layer hidden state to a delta spike input at $t=0$. In networks with diagonal $\mathbf{R}$ (Right), neurons act as leaky integrators that decay smoothly at different rates, mirroring biological \textit{temporal context cells}.
\textit{f. Temporal dynamics of the readout ($\mathbf{Lh}_t$)}. Response of the projected output neurons. The banded matrix structure in $\mathbf{L}$ (specific to SITH-RNN, \textit{Right}) generates translated eigenvectors, producing smooth, sequential activations with a spectrum of time constants that resemble biological \textit{time cells}.}
\label{fig:seqRNN}
\end{figure*}

\begin{figure}[!th]
\begin{center}
    \includegraphics[width=0.98\columnwidth]{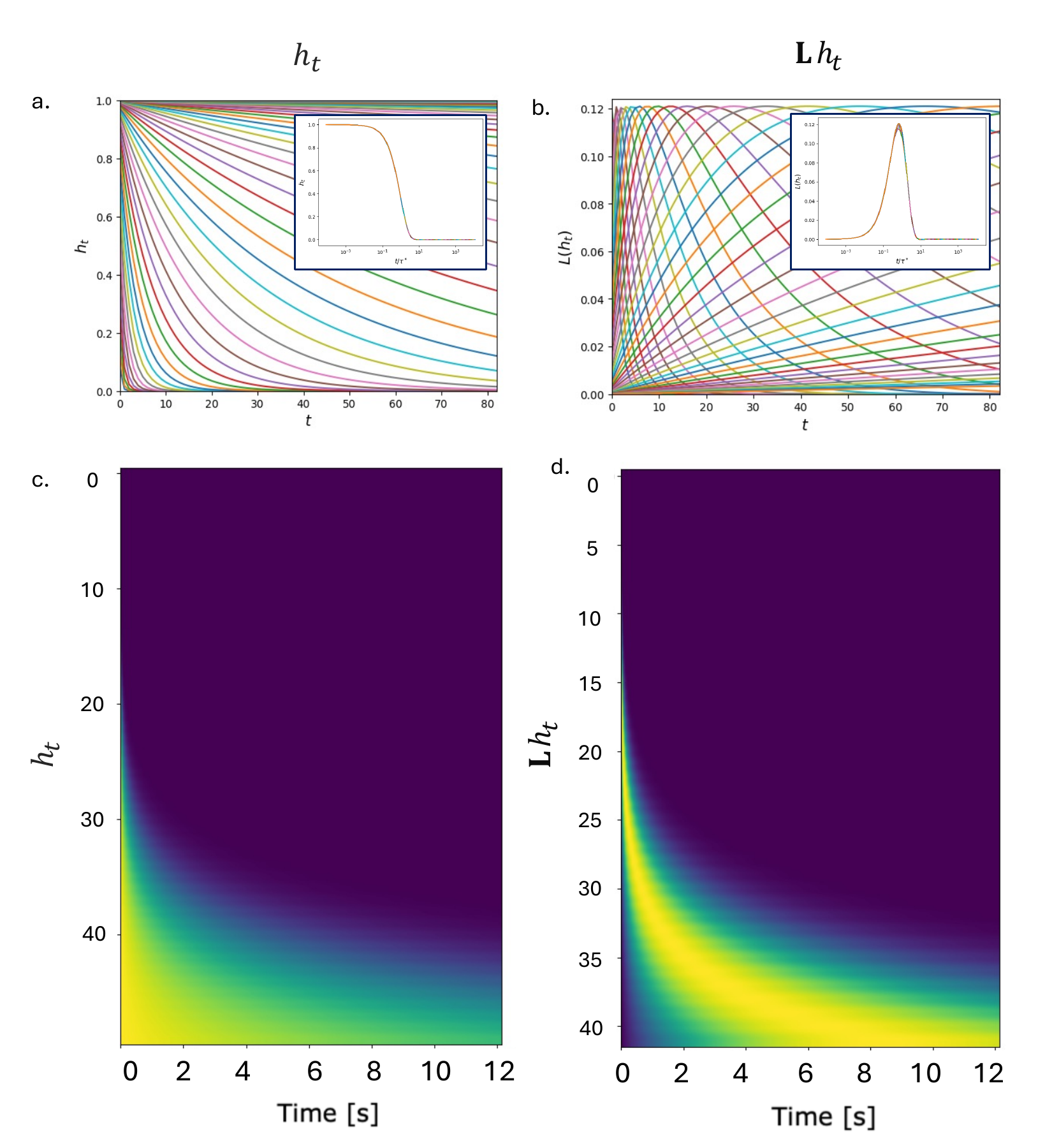}    
\end{center}
\caption{
\emph{SITH-RNN exhibits scale-invariant and sequential activity in the ``dual" hidden and projected output neuronal populations.} The temporal dynamics of the hidden layer $\mathbf{h_t}$ (\textit{Left}), and the feedforward output $\mathbf{Lh_t}$ (\textit{Right}), are shown corresponding to delta spike input for a single feature at time $t=0$, using individual neuron activation trajectories (\textit{Top}),  and a population heat map with neuron unit on the y-axis and time elapsed on the x-axis (\textit{Bottom}).
Insets in the top two figures show the activations when the x-axis is now rescaled by the time-constants of each neuron ($t/\tau_i$), which now line up onto each other.
The two different populations have markedly different features, with the hidden neurons all switching on at the same time and decaying at different rates (similar to temporal context cells), and the feed-forward neurons switching on sequentially after the spike with successively larger temporal fields (similar to time cells).
\label{fig:scale-inv}}
\end{figure}

To understand the underlying dynamics of SITH-RNN, we analyzed the impulse response of the network following a delta spike input. In Fig.~\ref{fig:seqRNN}e,f, we track the progression of time-dynamics for the hidden state ($\mathbf{h}_t$) and the projected output ($\mathbf{Lh}_t$) across the sequence of RNN architectures. We observe that constraining the recurrent weights $\mathbf{R}$ to be diagonal, with either uniform (in \textit{Network 3}) or geometric eigenvalues (in \textit{Network 4}) turns the hidden layer into a bank of integrators where neurons decay smoothly at different rates. 

However, sequential activity in the feedforward dynamics is only seen in SITH-RNN (\textit{Network 5}) which features a second constraint: the banded matrix structure in the readout $\mathbf{L}$---this generates translated eigenvectors, transforming the decaying hidden states into smooth, sequentially activated peaks in the readout $\mathbf{Lh}_t$ (Fig.~\ref{fig:seqRNN}f, right). Unlike the fixed analytic weights in SITHCon (which approximate an inverse Laplace transform), the motifs in $\mathbf{L}$ here are fully trainable; the only constraint is a zero-mean structure, which is sufficient to render sequential fields with increasingly wide receptive fields.

These activations are shown to be truly scale-invariant in Fig.~\ref{fig:scale-inv}. When the activation profiles of $\mathbf{h}_t$ and $\mathbf{Lh}_t$ are replotted as a function of normalized time ($t/\tau_i$, where $\tau_i$ is the time constant of that neuron), the curves collapse onto a single universal function (inset), satisfying the condition for scale-invariance in Eq.~\ref{eq: scale-inv}. We confirmed via ablation that this property is strict: scale-invariance emerges if and only if the network possesses \textit{both} geometric recurrent eigenvalues \textit{and} translated motifs in the readout (Fig.~\ref{fig:scale-inv2}).

Finally, the population heatmaps in Fig.~\ref{fig:scale-inv} reveal a direct isomorphism with biological memory circuits. The two distinct populations in the model map onto known neural phenotypes: the hidden neurons ($\mathbf{h}_t$), which switch on simultaneously and decay at diverse rates, replicate the dynamics of temporal context cells; meanwhile, the readout neurons ($\mathbf{Lh}_t$), which activate sequentially with widening fields, replicate the dynamics of time cells. This confirms that SITH-RNN does not merely solve the task but does so by instantiating a biologically plausible, scale-invariant temporal basis.

\subsection{Scale-invariant priors enable zero-shot generalization to time-rescaling}

To quantify the benefits of these inductive priors, we trained the networks on the hierarchical classification task introduced in Section 2. This task requires the model to implicitly learn the boundaries between letters, words, and sentences to predict the next token. We trained all models on sequences at a standard timescale ($3^0$). We then tested their ability to generalize zero-shot to sequences rescaled by factors up to $3^6$ (from $1x$ to $729x$ slower).

As shown in Fig.~\ref{fig:seqRNN}c, the generic linear RNN successfully learns the task at the training scale ($3^0$), demonstrating that a simple linear recurrence possesses sufficient memory capacity to solve the task. However, its performance declines immediately as the timescale shifts. We observe a clear trend across the architectural continuum: models perform increasingly well from left to right as each inductive prior is introduced. On the right, SITH-RNN achieves perfect classification accuracy over all \textit{six} orders of magnitude of the testing scale. This dissociation highlights that while generic linear dynamics can \textit{memorize} a specific temporal pattern, only the geometric inductive bias allows the network to \textit{generalize} the relative hierarchical structure independent of absolute duration.

\begin{figure}[t]
\begin{center}
    \includegraphics[width=0.95\columnwidth]{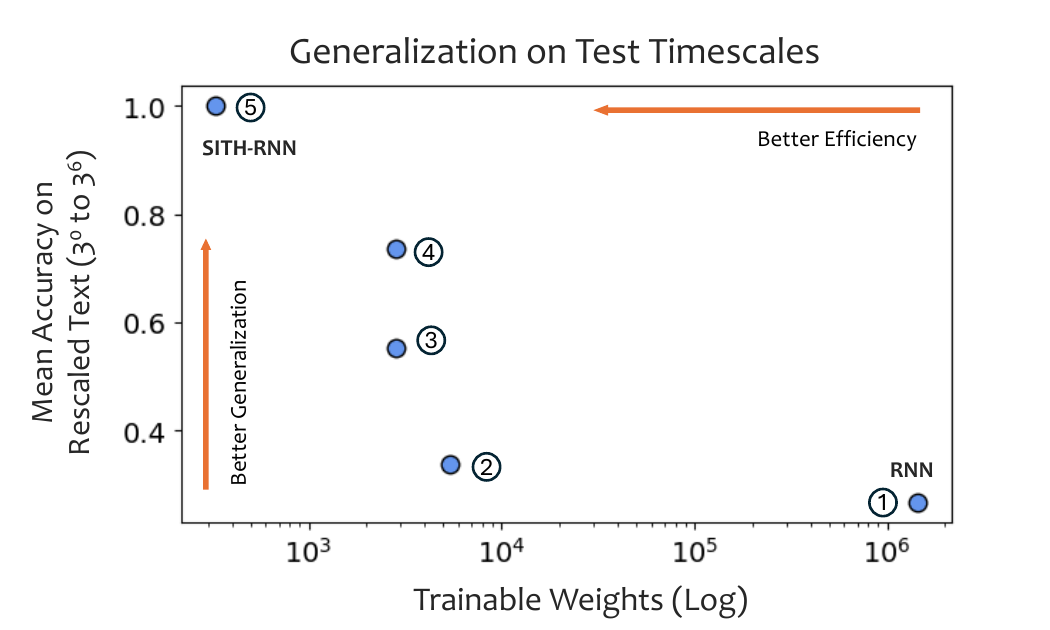}     
\end{center}
\caption{
\textit{Scale-invariant inductive priors enable robust generalization with minimal model complexity.}
We plotted the zero-shot generalization performance (averaged classification accuracy across six orders of magnitude of timescale, from Fig.~\ref{fig:seqRNN}c) against the number of trainable parameters for each architecture in the continuum.
Generic RNNs (\textit{Network 1}) exhibit high model complexity but fail to generalize to out-of-distribution timescales.
By contrast, imposing SITH-based inductive priors (\textit{Networks 2--5}) enables a simultaneous reduction in parameter count and improvement in robustness.
The final SITH-RNN architecture (\textit{Network 5}) optimizes this efficiency-generalization relationship, achieving perfect zero-shot generalization with significantly fewer trainable weights ($<0.05\%$ of the generic RNN).
\label{fig:genvweights}}
\end{figure}

\section{Discussion}

\begin{figure*}[t]
\begin{center}
    \includegraphics[width=0.85\textwidth]{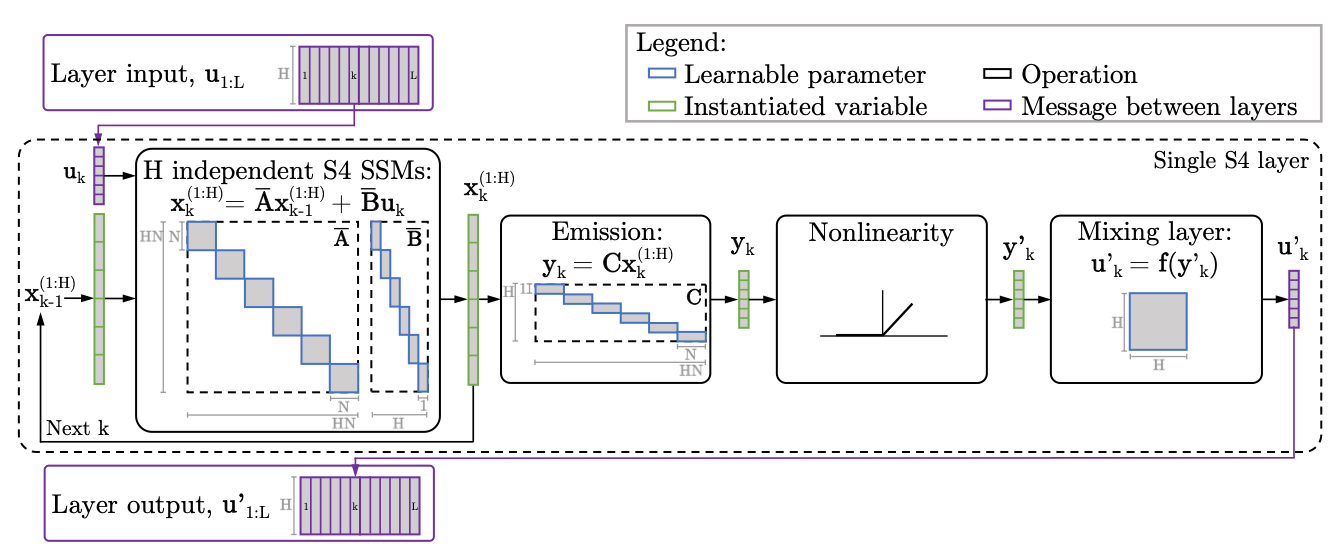}    
\end{center}
\caption{\emph{State-space models (SSMs) can be visualized as a block-diagonal system, similar to SITH-RNN, but with independent weights for each feature.} The schematic visualizes the dynamics for S4 \protect\cite{GuEtal21}, as it processes an input $\mathbf{u}$ with $H$ features and $L$ time steps.
At a single time-step, each input feature is passed along to $H$ independent SSMs, which expand on it and create a state space spanning $N$ dimensions---generating a memory $\mathbf{x}$ with dimensions $HN$.
This can also be visualized as a combined, single recurrent network with a block-diagonal recurrence matrix $\mathbf{\bar{A}}$ of shape $HN\times HN$, and block-diagonal projection weights take the input vector $\mathbf{u}_k$ and expand each feature into its state space ($\mathbf{\bar{B}}$), and project the $HN$-dimensional state vector $\mathbf{x}_k$ back into a $H$-dimensional output $\mathbf{y}_k$.
Both the recurrence and projection weights are block-diagonal, so the state space for each feature remains unaffected by the other features (save for a mixing layer at the end), similar to SITH-RNN, although the sub-blocks for each feature are independent in such SSMs.
Figure taken from \protect\citeA{SmitEtal22}.}
\label{fig:ssm}
\end{figure*}

The results presented here offer a resolution to the apparent paradox between local circuit heterogeneity and global cortical hierarchy. By validating a specific set of inductive priors---which create independent, logarithmically-compressed timelines for each feature, and scale-invariant dynamics, in both feedforward and recurrent architectures, we identify a candidate mechanism for how the brain constructs narrative structure from sensory streams. In this discussion, we examine the theoretical implications of this mechanism, propose falsifiable predictions for electrophysiology, and situate our findings within the broader context of temporal modeling in machine learning.

\subsection{The emergence of a discrete processing hierarchy from local temporal heterogeneity}

The organization of the human cortex into a functional hierarchy of increasing Temporal Receptive Windows (TRWs) suggests that the brain processes information through discrete, cascading timescales. However, the circuit mechanisms supporting this global architecture have remained elusive, particularly given the heterogeneity of time constants observed within local circuits. Here, we posit that such a heterogeneity of local time constants can not only co-exist, but also spontaneously give rise to, an increasing hierarchy of processing timescales.

We validated this using SITHCon, a deep feedforward network where each layer possesses an \textit{identical} distribution of intrinsic time constants but is constrained to represent a compressed memory of ``what" happened ``when". Despite the lack of an intrinsic gradient in cellular timescales, a functional hierarchy of TRWs emerged spontaneously during training. When input sequences were scrambled at different granularities, early layers were sensitive only to fine-grained perturbations (e.g., swapping letters), whereas deeper layers were sensitive not only to local jitter, but also when broad narrative structures (e.g., \textit{`paragraphs'}) were disrupted, mirroring the functional organization observed in human neuroimaging studies \cite{HassEtal08,LernEtal11}. Consistent with the principles of Slow Feature Analysis \cite{WiskSejn02}, this finding identifies local temporal heterogeneity as the necessary mathematical substrate for `renormalizing' fast sensory streams into stable, abstract narrative structures.

\subsection{Inductive priors for scale-invariance in recurrent neural networks}

While feedforward architectures can capture these principles, the brain must implement them through recurrent dynamics. We systematically introduced inductive priors inspired by the neuroscience of memory into a Recurrent Neural Network framework.
First, we enforced a separable representation by committing to a product hidden state with separate temporal spaces for each feature. This requires block-diagonal weight matrices where the recurrence can be factored as $\mathbf{R}_{\tt{what}}\otimes\mathbf{R}_{\tt{when}}$, creating a modular memory of feature history.

However, we found that modularity alone is insufficient for robust temporal processing. By analyzing the network as a dynamical system with recurrent weights $\mathbf{R}$ and downstream readout weights $\mathbf{L}$, we demonstrate that the network generates scale-invariant sequential activity only under strict spectral constraints: $\mathbf{R}$ must possess \textit{degenerate geometrically spaced} eigenvalues (implementing log-compression), and $\mathbf{L}$ must be structured with \textit{translated motifs} (implementing translation-equivariance) \cite{LiuHowa20}.

The resulting architecture, SITH-RNN, exhibits dual population dynamics resembling biological \textit{temporal context cells} (in the hidden state) and \textit{time cells} (in the readout). These constraints grant the network the ability to generalize zero-shot to out-of-distribution timescales---a feat unattainable by generic RNNs---while requiring dramatically fewer trainable weights. This confirms that the specific geometric timescale distribution of biological time cells is a normative solution for scale-invariant computation.

\subsection{Dissociating intrinsic dynamics from functional processing timescales}

How does the brain bridge the gap between millisecond-level cellular processes and the slow unfolding of narrative structure? In this paper, we argue that this hierarchy arises from network architecture rather than only single-neuron properties, and that the emergence of functional hierarchy depends critically on the topological organization of the readout connectivity rather than a slowing of intrinsic clocks.

\subsubsection{Experimental dissociation of intrinsic and emergent timescales}
In a naturally optimized biological circuit (or a standard deployment of SITH-RNN), one would typically expect the distribution of intrinsic time constants to scale with hierarchy, with deeper layers accessing progressively longer $\tau_{max}$ \cite{JacqEtal21,JacqEtal22}. However, in this study, we intentionally restricted our model to possess \textit{identical} distributions of intrinsic time constants across all layers. This experimental control was critical: it ensures that the observed hierarchy of TRWs is not merely \textit{hard-coded} by the cellular properties, but must instead arise as an emergent functional property of the network architecture. Indeed, despite this restriction, deep layers still emerged as \textit{`sentence'}-level and \textit{`paragraph'}-level processors.

\subsubsection{Biophysical agnosticism of slow dynamics}
These findings challenge the view that the cortical hierarchy of Temporal Receptive Windows (TRWs) is strictly dictated by a gradient of intrinsic cellular properties, such as receptor kinetics or membrane time constants \cite{MurrEtal14, ChauEtal15}. While such anatomical gradients are well-documented \cite{SiegEtal21}, our results suggest that they might not be sufficient to explain the full magnitude of functional separation observed in the brain. Indeed, recent electrophysiological evidence from the ferret auditory cortex confirms that integration windows are organized hierarchically across regions—driven by circuit topology—rather than cortical layers, suggesting that local laminar gradients alone cannot account for the emergence of narrative timescales \cite{SabaEtal25}.

Furthermore, the mathematical principles derived here are agnostic to the specific biophysical source of these timescales. In our model, slow dynamics arise from the eigenvalues of the recurrent matrix $\mathbf{R}$. However, in biological circuits, such dynamics need not rely solely on reverberating loops; they can also emerge from intrinsic cellular mechanisms, such as calcium-activated nonspecific cation currents \cite{FranEtal02,LiuEtal19} or slow synaptic kinetics \cite{GuoEtal21}. Indeed, the uncoupled, diagonal structure of SITH-RNN is functionally equivalent to a bank of parallel neurons driven by intracellular processes.

\subsubsection{Topological organization drives temporal renormalization}
The emergence of this hierarchy suggests that the transition from sensation to narrative is driven by a structural renormalization of the timeline rather than just a slowing of single-neuron clocks. By max-pooling over temporal motifs, deep layers explicitly discard the precise timestamp of an event in favor of its semantic identity. Crucially, this implies that the generation of slow timescales alone is insufficient; the critical constraint is the \textit{organization} of the readout, which effectively implements the translation-equivariant motifs described in Section 4.2. Indeed, shared or redundant functional connectivity is essential for robust temporal processing \cite{MachEtal10}. Consequently, anatomical connections in the brain cannot be unstructured; they must respect the topology of the temporal memory, ensuring continuity across timescales. Whether maintained via active firing in continuous attractor networks \cite{DaniHowa25,SarkEtal24} or \textit{activity-silent} synaptic traces \cite{Stok15,MongEtal08}, the system must adhere to these geometric constraints to support the renormalization of sensory information into narrative structure.

\subsection{Convergent mechanisms and divergent scaling laws in biological versus machine memory}

There is a growing convergence between the mechanisms used by the brain to store temporal history and modern techniques for sequence modeling in AI. In this section, we trace the evolution of temporal basis functions in machine learning---from early recurrent architectures to modern State Space Models. We highlight their structural convergence with, yet spectral divergence from, biological memory circuits.

\subsubsection{Evolution of temporal basis functions in machine learning}
Recurrent neural networks have long struggled to model long-range temporal dependencies, primarily due to the vanishing gradient problem. While architectures like Long Short-Term Memory (LSTM) networks mitigated this issue \cite{HochSchm97}, they lack biological plausibility and often fall short when capturing dependencies across very long timescales. 

To address this, machine learning research has increasingly gravitated toward models that project history onto a fixed, high-dimensional temporal basis—an approach structurally reminiscent of Reservoir Computing \cite{MaasEtal02,Jaeg01}. Notably, Legendre Memory Units (LMUs) \cite{VoelEtal19} and the HiPPo framework \cite{FuEtal22} attempt to approximate function history using orthogonal polynomial bases.

This polynomial projection approach evolved into the broad family of State Space Models (SSMs; see \citeNP{PatrAgne24}), governed by the general equations:
\begin{eqnarray}
    \mathbf{h}^{SSM}_{t} = \mathbf{A} \mathbf{h}^{SSM}_{t-1} + \mathbf{B}x_t\;, \label{eq:ssm} 
    \\ y_t =  \mathbf{C}\mathbf{h}^{SSM}_t \label{eq:ssm_o_t}
\end{eqnarray}
where $\mathbf{A}$ evolves the hidden state and $\mathbf{B, C}$ are projection operators. Critically, these models typically instantiate independent state-space dynamics for each feature, creating a structure that mirrors the block-diagonal priors we propose (see Fig.~\ref{fig:ssm}).

\subsubsection{Shared diagonal architecture, distinct temporal bases}
There is a striking formal alignment between the biologically constrained SITH-RNN and modern SSMs like S4D \cite{GuEtal22} or Mamba \cite{GuDao23}. Both architectures solve the problem of temporal integration by diagonalizing the recurrence matrix, effectively decomposing complex histories into independent, orthogonal memory channels.\footnote{The evolution equations for SITH-RNN (Eqns. \ref{eq:rnn}--\ref{eq:o_t}) are formally analogous to the discretized SSM formulation. Specifically, the SSM recurrence matrix $\mathbf{A}$ maps to our diagonal decay matrix $\exp(-\mathbf{s}\Delta)$; the input projection $\mathbf{B}$ maps to our weights $\mathbf{I}$ (scaled by decay); and the output $\mathbf{C}$ maps to our readout $\mathbf{L}$.} This suggests that the decoupling of temporal modes is a fundamental computational principle shared by biological evolution and machine learning engineering.

However, while the \textit{mechanisms} have converged, with recent models like S4D \cite{GuEtal22} and Mamba \cite{GuDao23} moving toward the diagonal initializations inherent to SITH-RNN---the \textit{spectra} remain fundamentally different. Standard SSMs prioritize polynomial bases (e.g., Legendre polynomials in HiPPo) for optimal memory reconstruction, but these are mathematically tethered to fixed interval scales \cite{GuEtal21}. In contrast, biological circuits utilize a \emph{scale-invariant} geometric eigenspectrum. This establishes a logarithmic ratio scale—consistent with the Weber-Fechner law, which describes the logarithmic relationship between physical stimulus magnitude and internal sensation \cite{Fech60}—rather than a linear one. While recent innovations in Mamba offer improved expressivity, they lack this inductive prior and thus do not inherently support zero-shot timescale generalization.

\subsection{Toward robust, scale-invariant machine intelligence}

We posit that the Weber-Fechner law is not simply a biological feature, but a fundamental computational constraint required for robust sequence modeling. Specifically, logarithmic tiling allows memory horizons to grow exponentially with a linear increase in neurons, offering a geometric mechanism to bypass the context-length bottlenecks of current Large Language Models.

These context-length bottlenecks largely arise due to the quadratic complexity—across training, inference, and memory usage—inherent to the underlying self-attention architecture \cite{VaswEtal17}. Recent efforts have begun to operationalize the synergy between biological priors and machine learning to address this limitation. By either preprocessing the input history into a scale-invariant timeline \cite{DickTiga25} or embedding the memory timeline directly into the attention block to create a ``time-local" Transformer \cite{DickEtal25}, these approaches demonstrate that log-compressed SITH memory can extend the effective context capacity of modern AI without incurring the prohibitive costs of full self-attention.

Furthermore, this prior provides a principled solution for automatic speech recognition: unlike standard architectures that require extensive training data to handle variable speech rates (e.g., wav2vec; \citeNP{BaevEtal20}), scale-invariant dynamics naturally adapt to faster or slower inputs without retraining. Recent findings confirm that this computational capability mirrors the biological reality of the ferret auditory cortex, with neuronal integration windows that remain stable and context-invariant even as the information rate of speech varies \cite{SabaEtal25}. This suggests that the brain---similar to SITH-RNN---relies on a population code over fixed temporal filters rather than dynamically rescaling individual cellular clocks. Merging this biological constraint with the efficiency of modern state-space models could thus combine the computational scalability of machine learning with the temporal robustness of biological intelligence.

\section{Methods}
\begin{table*}[htbp]
\caption{Summary of Recurrent Network Architectures and Constraints}
\label{tab:rnn_constraints}
\centering
\begin{tabular}{lccc}
\toprule
\textit{Model} & \textit{Recurrence Matrix} $\mathbf{R}$ & \textit{Eigenvalues} $\lambda$ & \textit{Readout} $\mathbf{L}$ \\
\midrule
1. Generic RNN & Dense & Random & Dense \\
2. Block-Diagonal & Block-diagonal & Random & Dense \\
3. Diagonal (Uniform) & Block-diagonal & Uniformly Spaced & Dense \\
4. Diagonal (Geometric) & Block-Diagonal & Geometrically Spaced & Dense \\
5. SITH-RNN & Block-Diagonal & Geometrically Spaced & Banded (Toeplitz) \\
\bottomrule
\end{tabular}
\end{table*}

\subsection{Hierarchical Language Corpus}
We constructed a hierarchical dataset using a deterministic grammar to simulate the nested structure of natural language. The vocabulary consisted of symbols $\{1, \dots, 9\}$, representing the lowest hierarchical level (\textit{`letters'}). These symbols were recursively combined into higher-order structures: triplets of \textit{`letters'} formed \textit{`words'}, triplets of \textit{`words'} formed \textit{`sentences'}, and so on. This combination process relied on strided positions; for example, the symbols at the first, fourth, and seventh positions were combined to create the first entry ($147$) of the subsequent hierarchical level (\textit{`words'}). Similarly, the first, fourth and seventh words ($147$, $267$, and $357$ respectively) were combined to create the first entry in the next hierarchical level (\textit{`sentences'}) (see Fig.~\ref{fig:toylang}b).

Symbol transitions within triplets followed a deterministic, context-dependent grammar designed to emulate XOR logic: the identity of the third symbol was uniquely determined by the first two, but could not be predicted from either alone. To prevent positional clustering (e.g., the symbol $1$ consistently appearing at the start of a hierarchical unit) and ensure structural diversity, the symbol lists were randomly scrambled before being combined at each level.

Since these rules are recursive, they allow for the generation of an unbounded number of hierarchical levels, where symbol transitions at lower levels change abruptly when higher-level boundaries are reached. For this study, the final sequences comprised \textit{four} hierarchical levels with a total length of $3^4 = 81$ symbols. The network was trained to classify \textit{nine} possible input sequences to their corresponding labels. This task required the network to track symbols across multiple time-scales, akin to maintaining a recollection of words, paragraphs, and larger context required to understand speech

\subsection{SITHCon Architecture}
The SITHCon architecture \cite{JacqEtal22} served as the feedforward baseline. Temporal memory was generated via a bank of filters $f_i(t)$ that approximated a logarithmic timeline.
These filters were constructed using the inverse Laplace transform of the function $F(s) = s^{-k}$ (the Post approximation), where $k$ controls the temporal acuity.
We utilized $k=15$ for all experiments.
The memory bank consisted of $n_\tau = 50$ filters with time constants $\tau_i$ geometrically spaced between $\tau_{min}=1$ and $\tau_{max}=81$.

This setup creates a Scale-Invariant Temporal History (SITH) representation where a temporal rescaling of the input $x(t) \to x(\alpha t)$ results in a translation of the memory pattern along the neuron index $i$.
A dense convolutional layer then processed this memory representation.
Because the convolution operation is equivariant to translation, the network could learn features that were invariant to the absolute duration of the input patterns.
To facilitate comparison with RNNs, we removed the final ReLU nonlinearity and linear projection layer, restricting learning to the convolutional kernels.

\subsection{Network Architecture and Training}
We trained a sequence of recurrent neural networks (ranging from generic to SITH-RNN) and the feedforward SITHCon model.
Both recurrent and feedforward architectures were constructed with $4$ layers, to match the levels of hierarchy in the toy language.
Each layer contained $9$ input channels (features) and a temporal space of $n_{\tau}=50$ for each feature, resulting in a total hidden state dimension of $N_{\tt{hid}} = 9 \times 50 = 450$ dimensions.
For Networks 2 through 5 and SITHCon, the readout is followed by a convolutional layer with kernel width $g_{\tt{width}}=1$ (spanning a single time constant) and a maxpool operation to create remapped features for the next layer.

All models were implemented in PyTorch and trained on the hierarchical language classification task.
We utilized the \textit{AdamW} optimizer with a weight decay parameter of $wd = 0.001$ to prevent overfitting.
Training was conducted for $n_{\tt{epoch}}=200$ epochs.
Models were trained on sequences at the standard timescale ($1\times$), corresponding to the base symbol duration.

\subsection{Temporal Recurrence Analysis}
To quantify the temporal structure of layer representations, we computed self-similarity matrices for the output activity $\mathbf{o}^l_t$. The similarity between any two time points $t_i$ and $t_j$ was defined as the Pearson correlation coefficient between their population vectors:
\begin{equation}
    \rho_{\mathbf{o}(t_i),\mathbf{o}(t_j)} = \frac{(\mathbf{o}(t_i)-\langle \mathbf{o}(t_i)\rangle) \cdot (\mathbf{o}(t_j)-\langle \mathbf{o}(t_j)\rangle)}{\norm{\
    \mathbf{o}(t_i)-\langle\mathbf{o}(t_i)\rangle}{2}\;\norm{\mathbf{o}(t_j) - \langle\mathbf{o}(t_j)\rangle}{2}}  
\end{equation}
where $\langle \mathbf{o}(t) \rangle$ denotes the mean activity across neurons at time $t$. This metric is equivalent to the cosine similarity of the z-scored population vectors.

\subsection{Measuring sensitivity of layers to different timescales by permuting sequences}
To quantify the effective TRW of each layer, we applied a temporal scrambling perturbation analysis.
Input sequences $x$ were permuted at various granularities $s$, ranging from individual symbols ($3^0$) to entire sequences ($3^6$).
For this analysis, test sequences were constructed by concatenating 9 classification sequences to create a longer continuous stream of length $N=729$.
Let $P_s[x]$ denote the permuted sequence. We fed this permuted sequence into the network and recorded the layer activations $\mathbf{o}^l(P_s[x])$.

To compare these activations with the original response $\mathbf{o}^l(x)$, we essentially ``un-shuffled" the output by applying the inverse permutation $P_s^{-1}$ to the temporal dimension of the activations.
We then computed the correlation between the recurrence matrices ($\rho$) of the original and the un-shuffled responses:
$C(\rho^l, \rho^l_{\tt{shuff}}(s))$.
A high correlation indicates that the layer's representation is robust to scrambling at scale $s$, implying that its integration window is smaller than $s$.
Conversely, a drop in correlation indicates that the layer relies on temporal structure at that scale.
This analysis was repeated for $n_{\tt{shuff}}=200$ random permutations per scale to ensure statistical robustness.

\subsection{Sequence of Recurrent Networks}
We systematically constrained the generic RNN formulation (Eq.~\ref{eq:rnn}) to test specific biological priors. The architectural variants are summarized in Table 1.
Weights were shared across layers, mirroring the SITHCon architecture to isolate the effect of recurrence.

\textit{Generic RNN:} To ensure a direct architectural comparison with the linear SITH-RNN, the Generic RNN (Model 1) was implemented as a deep linear network. It possessed the same hidden dimension ($N_{hid}=450$) as the block-diagonal networks but utilized dense, fully connected recurrent matrices. Weights were initialized from a uniform distribution scaled by the inverse square root of the layer size (He initialization variant), and no element-wise nonlinearity was applied between time steps.

\paragraph{Implementation Details}

\subsubsection{Discretization of continuous-time equations}
We adapt the continuous-time constraints for scale-invariance from \citeA{LiuHowa20} for discrete-time simulation (step $\Delta t$) by modeling the hidden state evolution via the diagonal matrix $\mathbf{R}_{\text{when}} = \text{diag}\left(e^{-\lambda_{1}\Delta t}, \dots, e^{-\lambda_{n}\Delta t}\right)$. Furthermore, we model inputs as instantaneous pulses at the step onset rather than persistent signals. Consequently, the input projection is scaled to reflect the heterogeneous decay across time constants over the interval:

\[
  \mathbf{I}_{when} =
  \begin{bmatrix}
    \exp(-\lambda_{1}\Delta t) \cdots \exp(-\lambda_{n}\Delta t)
  \end{bmatrix}.
\]

\subsubsection{Distribution of Eigenvalues in $\mathbf{R}$}

For the diagonal models (3-5), each layer had the same distribution of time constants. For the Uniform model (3), eigenvalues were distributed as $\lambda_i = 1/(\tau_{min} + \alpha i)$, enabling linear spacing of time constants.
For the Geometric models (4 and 5), the time constants $\tau_i = 1/\lambda_i$ were distributed according to:
\[
    \tau_i = \tau_{min} \left( \frac{\tau_{max}}{\tau_{min}} \right)^{\frac{i}{n_\tau - 1}}
\]
which ensured that the $n_{\tau}$ neurons tile $\tau_{min}=1$ to $\tau_{max}=81$ evenly in log time.

where $\tau_{min}$ is the minimal time constant, and $c$ is chosen such that the $n_{\tau}$ neurons tile $\tau_{min}$ to $\tau_{max}$ evenly in log time. Each layer has the same distribution of time constants. 

\subsubsection{Specific constraints on the readout matrix $\mathbf{L}$}
To implement the translation-invariant eigenvector constraint from \citeA{LiuHowa20}, we distinguish the \textit{SITH-RNN} from previous block-diagonal implementations by restricting the form of $\mathbf{L}_{\text{when}}$. Rather than being fully trainable, $\mathbf{L}_{\text{when}}$ is confined to a banded, Toeplitz-like structure, ensuring the effective recurrent matrix supports translated eigenvectors.

We construct $\mathbf{L}_{\text{when}}$ using a repeating motif that slides along the diagonal. The width of this motif, $L_{\text{width}}$, is a hyperparameter kept small relative to the number of time constants $n_{\tau}$. The values within the motif are trainable, shared across layers, and constrained to sum to zero. For example, with $L_{\text{width}} = 3$ and $n_{\tau}= 6$, the matrix takes the form:
\[
  \mathbf{L}_{\text{when}} =
  \begin{bmatrix}
    b & c & & & & \\
    a & b & c & & & \\
    & a & b & c & &\\
    & & a & b & c & \\
    & & & a & b & c \\
    & & & & a & b 
  \end{bmatrix}.
\]
In the specific case of \textit{SITH-RNN} reported here, we employ a fully trainable motif with fixed width $L_{\text{width}} = 7$.

\section*{Author Note}
This work was supported by a Rajen Kilachand Fund award to Michael Hasselmo and NIMH R01MH132171 to MWH. 
MWH is a co-founder of Cognitive Scientific AI, which may benefit indirectly from the publication of this study.

\bibliography{bibdesk}

\clearpage
\onecolumn
\section{Supporting Information (SI)}
\setcounter{figure}{0}
\renewcommand{\figurename}{Fig.}
\renewcommand{\thefigure}{S\arabic{figure}}

\begin{figure}[!h]
\begin{center}
    \includegraphics[width=0.78\textwidth]{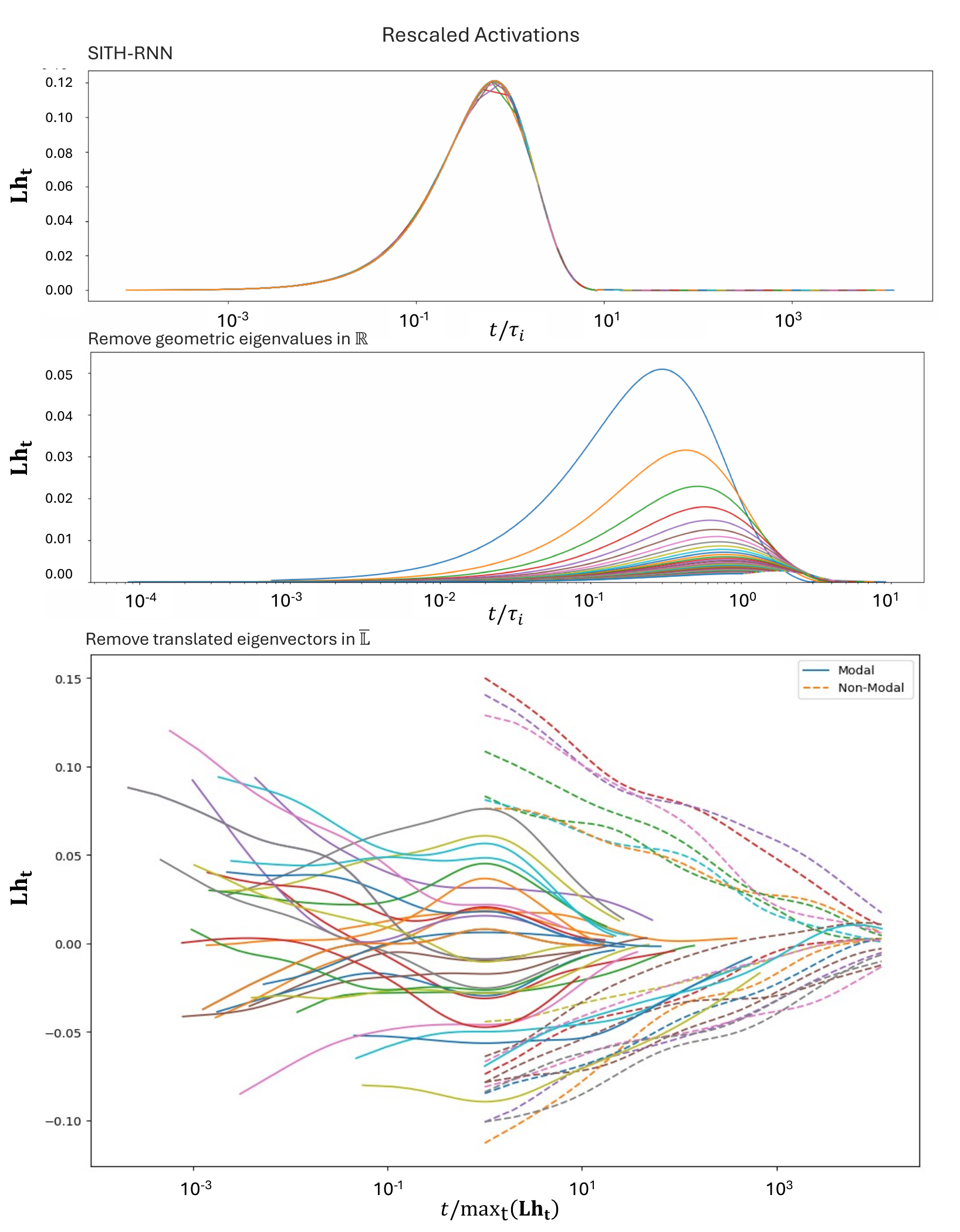}    
\end{center}

\caption[Ablation study confirms that scale-invariance requires the conjunction of geometric eigenvalues and translation-equivariant readouts.]{\emph{Ablation study confirms that scale-invariance requires the conjunction of geometric eigenvalues and translation-equivariant readouts.} 
We analyzed the impulse response of the output population $\mathbf{Lh}_t$ to a delta input at $t=0$ across three model variations.
\emph{Top (SITH-RNN):} The full model, featuring both geometric eigenvalues in $\mathbf{R}$ and banded, translation-equivariant readouts in $\mathbf{L}$. When plotted against rescaled time ($t/\tau_i$), the response curves of different neurons collapse onto a single, universal function, indicating true scale-invariance.
\emph{Middle (Uniform Eigenvalues):} Replacing the geometric distribution in $\mathbf{R}$ with a uniform distribution preserves sequentiality but destroys scale-invariance. The curves fail to align when rescaled, confirming that logarithmic compression is mathematically necessary for the scaling property.
\emph{Bottom (Dense Readout):} Preserving geometric eigenvalues but replacing the banded $\mathbf{L}$ with a standard dense, trainable matrix results in disordered dynamics. The population splits into ``Modal" (peaked) and ``Non-Modal" (monotonic) responses, losing the interpretable ``time cell"-like sequences entirely. Thus, both the geometric eigenvalues in  $\mathbf{R}_{when}$ and translated motifs in $\mathbf{L}_{when}$ are required in SITH-RNN to produce scale-invariant, sequential dynamics.
}
\label{fig:scale-inv2}

\end{figure}

\begin{figure*}[t]
\begin{center}
    \includegraphics[width=1.0\textwidth]{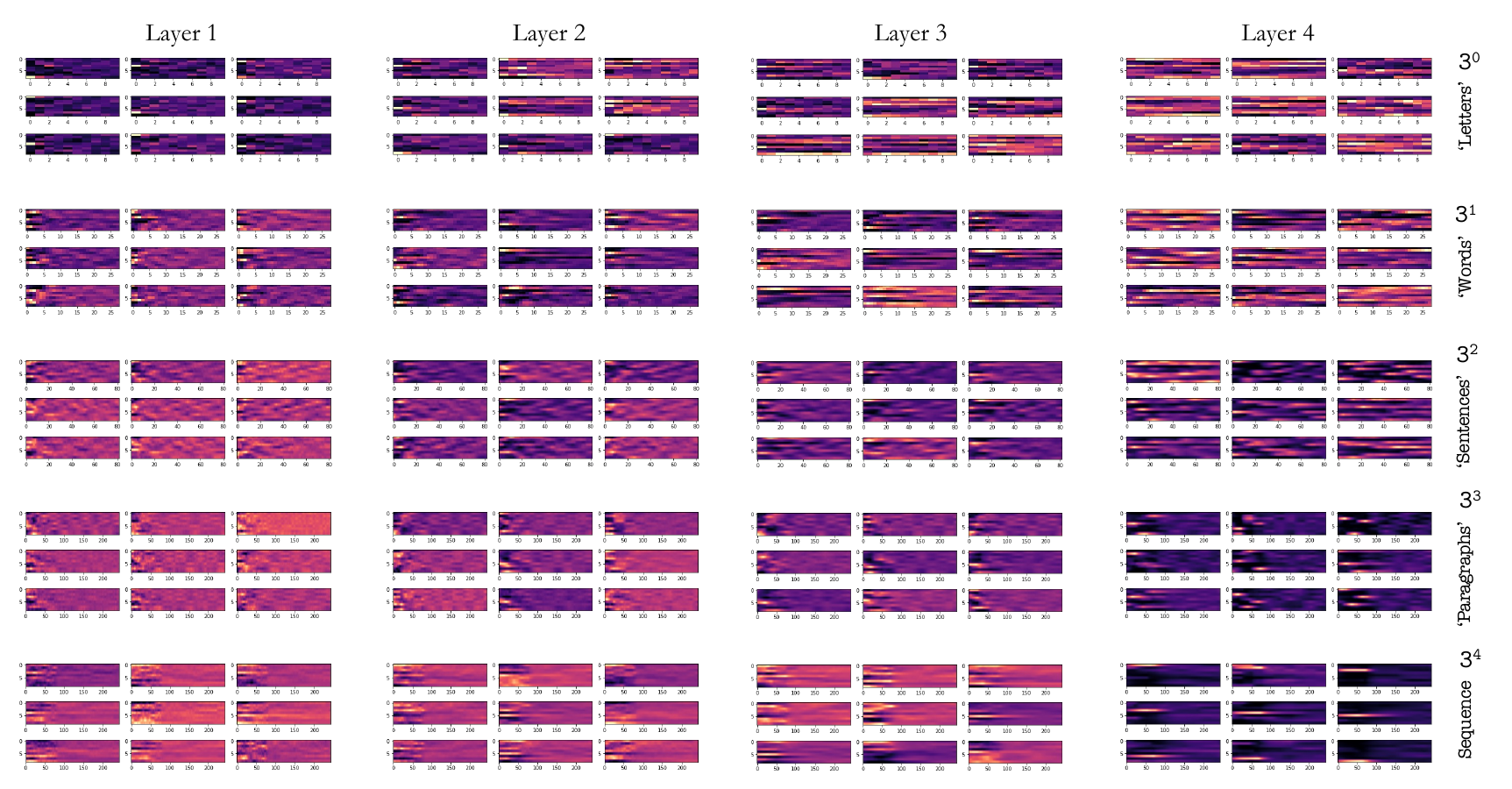}    
\end{center}

\caption{\textit{A comprehensive mapping of linear receptive fields via Spike-Triggered Averaging reveals layer-wise renormalization and learned compositional hierarchy.} 
    This figure expands upon the representative examples in Figure 4d by visualizing the linear receptive fields for the complete population of nine neurons across all four hidden layers. Receptive fields were recovered using Spike-Triggered Averaging (STA) relative to the onset of symbols at five hierarchical scales, ranging from elementary \textit{`letters'} ($3^0$) to full \textit{`sequences'} ($3^4$). The visualization highlights a fundamental transformation in temporal processing depth. \textit{Left:} Layer 1 neurons display sharp, diagonal banding at the finest timescales, indicating they function as precise detectors for local symbolic transitions---but fail to integrate global context. At higher levels (e.g., \textit{`sequence'} or \textit{`paragraphs'}), their receptive fields appear as dense, repetitive tiling patterns---showing a periodicity which indicates that early layers effectively ``reset'' with each local symbol transition, tracking recurring constituent elements (e.g., every instance of a specific letter triplet) without the capacity to distinguish their unique position within the broader narrative structure. \textit{Right:} In contrast, Layer 4 neurons (right column) exhibit a nested structure that mirrors the compositional rules of the grammar. Proceeding from the bottom to the top of the Layer 4 column, a single broad activation band at the \textit{`sequence'} level resolves into three distinct bands at the \textit{`paragraph'} level, which further subdivide into nine bands at the \textit{`sentence'} level. This fractal-like fractionation confirms that deep neurons have successfully learned the mapping between hierarchical levels, defining their receptive fields through the appropriate combinatorics of constituent symbols (e.g., a specific sequence is recognized as a composition of three specific paragraphs). \textit{Middle:} The transition between these extremes is not binary. The intermediate columns (Layers 2 and 3) exhibit a rich, continuous gradient of integration windows, which demonstrate the network's capacity for mixed selectivity, processing intermediate linguistic structures (e.g., phrases) that exist between the speed of a single symbol and the duration of a full narrative.}
\label{fig:STAsupp}

\end{figure*}

\end{document}